\documentclass[%
 aip,
 amsmath,amssymb,
 reprint,%
 floatfix,%
 raggedbottom,%
]{revtex4-1}

\usepackage{dcolumn}% Align table columns on decimal point
\usepackage{bm}% bold math
\usepackage[utf8]{inputenc}
\usepackage[T1]{fontenc}
\usepackage{mathptmx}

\usepackage{lipsum}% http://ctan.org/pkg/lipsum
\usepackage{graphicx}% http://ctan.org/pkg/graphicx
\usepackage{xcolor}
\makeatletter

\long\def\@makecaption#1#2{%
  \par\vskip\abovecaptionskip
  \begingroup
   \small\rmfamily\samepage\flushing
   \let\footnote\@footnotemark@gobble
   \noindent\@make@capt@title{#1}{#2}\par
  \endgroup
  \vskip\belowcaptionskip}
\makeatother
\usepackage{xspace}
\begin{document}

\title{Periodic and quasiperiodic structures produced by microphase separation in one-component simple liquids with competing interactions}

\author{Lorenzo Agosta}
\affiliation{Dept. of Chemistry, \AA ngstr\"{o}m laboratoriet, Uppsala Universitet, 751 21 Uppsala, Sweden}
\email{lorenzo.agosta@kemi.uu.se}
\author{Alfredo Metere}
\affiliation{Metere Industries, Inc., 350 Sansome St Ste 820 \#1055, San Francisco, California 94104, USA}
\author{Mikhail Dzugutov}
\affiliation{Dept. of Chemistry, \AA ngstr\"{o}m laboratoriet, Uppsala Universitet, 751 21 Uppsala, Sweden}

\date{13 November 2026}% date of submission to JCP

\begin{abstract}
  Competing interactions can generate frustration which suppresses
  macroscopic-scale phase separation, superseding it with separation
  patterns on the microscopic scale. This phenomenon, referred to as
  {\it microphase separation}, produces thermodynamically stable
  mesoscopic-scale periodic patterns in systems of different physical
  nature. The studies of this effect
  follow two paradigms.
  First, in block copolymers, the
  macroscopic-scale separation of chemically distinct components is
  prevented by the covalent bonds between them. In the other,
  the microscopic-scale separation is produced in simple
  single-component liquids by the competition between a
  short-range attraction and a long-range repulsion (SALR)
  in the pair potential, the separation occurring between a
  dense phase and a gas. The interest in the simple liquid
  models with SALR interaction was inspired by the apparent
  universality of the modulated structures generated by competing
  interactions in different systems. Simple particle models
  are tractable by theory and thereby make it possible to discern the
  microscopic mechanisms of pattern formation; particle-based
  simulations give access to systems beyond the reach of experiments. In this Perspective we review the recent
  advances in the molecular-dynamics simulation of microphase
  separation in single-component simple liquids using SALR potentials. We show that these simulations reproduce the
  mesophases typical of block copolymers and liquid crystals, which
  demonstrates that the complexity of the constituent particles of real
  liquids is not a necessary condition for their formation, and that
  they produce novel periodic morphologies of both conceptual interest
  and technological relevance; we discuss how simulated mesophases can
  serve as templates for metamaterials whose properties are governed by
  topology.
\end{abstract}

\maketitle

\section{Introduction}

The problem of general description of liquids still remains an
intriguing challenge for statistical mechanics. The central idea of
classical statistical mechanics is to use many-body mechanical models
for calculating the macroscopic properties of physical systems. The
behaviour of a mechanical model of $N$ particles is uniquely
determined by its Hamiltonian which is defined by the potential energy
describing the interaction of constituent particles. The ultimate goal
of the statistical mechanics of liquids is to establish a link between
the assumed interparticle interaction forces and the macroscopically
measurable properties, first of all the structure and the phase
behaviour. We are still far from solving this problem.

The basic difficulty in building a unifying description of the liquid
state is the absence of a reference ground model. For finite density
gases such a model is the perfect gas, which can be computed as statistics
of independently moving particles. This reference model can be
employed for the description of the macroscopic behaviour of a real
gas considering the gas density as a small parameter for free-energy
expansion. For crystalline solids the ground state model is the
particles' arrangement in a perfect periodic pattern. The amplitude of
small particle excursions relative to their perfect positions can be
used as a small parameter for calculating the macroscopic properties
of real crystals.  All the attempts to find an appropriate universal
ground-state model in liquids have been
unsuccessful.\cite{lebowitz_waisman_1980,trachenko_collective_2016} The
attempts to employ in this capacity the system of hard spheres using
both the dense random packing procedure and the integral equation
approach failed.\cite{rosenfeld_1979,parisi_slanina_2000} The ground state ``inherent structures'' of simple liquids\cite{weber_stillinger_1984,stillinger_weber_1985,dzugutov_icosahedral_1992}
were found to be strongly non-universal. Moreover, the inter-particle
interactions in liquids are strongly system-dependent. All these
reasons make it impossible to find a zero approximation as in the case
of gases or crystalline solids.  Liquids have neither the small
displacements of solids nor the small interactions of
gases. Summarized aptly by Landau, liquids have no small parameter.\cite{landau}

The Van der Waals theory was the first attempt to explain the
ubiquitous phenomenon of the gas-liquid condensation in simple systems
by linking it to a particular form of inter-particle interaction
potential.\cite{lebowitz_waisman_1980,johnston_2014}  It elucidated the physical effect of attractive
and repulsive forces relating the underlying mean-field approximation
to the force range. The van der Waals model assumes that
inter-particle pair interaction potential includes a steep short-range
repulsion part and a less steep attraction at longer separation
distances. The main result of that model was the prediction of a
distinct domain in the system's phase diagram where the system is
decomposed into two macroscopically separated fluid phases,
low-density gas and dense liquid, which coexist in equilibrium.

However, it was found that the macroscopic-scale phase separation
predicted by the Van der Waals model, called spinodal decomposition, is not a
universal feature of simple
systems.\cite{sear_gelbart_1999,archer_phase_2007} There are simple
systems lacking the spinodal domain in their phase diagrams, in which
the macroscopic coexistence is replaced by modulated
phases.\cite{malescio_pellicane_2003,imperio_reatto_2006,camp_2003,zhuang_prl_2016}

More than 50 years ago Lebowitz and Penrose
\cite{lebowitz_rigorous_1966} demonstrated that if the Van der Waals pair
potential is amended by an additional repulsive part of sufficiently
long range, the macroscopic decomposition into the distinct vapor and
liquid phases is supplanted by {\it microphase separation} where these
phases are separated on the scale of several particle diameters. Such
a potential is named SALR (short attraction long repulsion). The same
conclusion was reached in the language of Landau theory by
Brazovskii\cite{brazovskii_1975} for any system whose free energy is
minimised by density modulations of finite wave vector, and the
phenomenology of the resulting modulated phases, common to systems as
different as Langmuir monolayers, ferrofluids and block copolymers,
was reviewed by Seul and Andelman.\cite{seul_domain_1995}

Microphase separation transition was found in polymeric systems
resulting in the formation of structures with mesoscale periodic modulation of
different dimensionality which are disordered at the molecular scale. These include
lamellar and columnar periodic phases as well as lattices of droplets. The most intriguing possibility
is the formation of continuous triply periodic mesophases with a cubic
symmetry (gyroid) or other forms of 3D symmetry. This phase behaviour
has been independently observed in complex fluids like block
copolymers and liquid
crystals,\cite{leibler_theory_1980,Epps2004,meuler_ordered_2009,bates_network_2005}
and more recently also in globular protein--polymer
conjugates.\cite{lam_olsen_2013,lam_olsen_2014}

A conceptually interesting aspect of exploring these phases in simple
systems concerns the thermodynamic mechanisms of their stabilisation
and self-assembly. Since the seminal work by Onsager
\cite{onsager_notitle_1949}, the main idea of the mesophase formation
was based on the entropic paradigm, whereby the particle geometry played a
crucial role. It is therefore of interest to investigate the
possibility that these phases can be formed by particles with
spherically symmetric interaction. Experimentally, the suitable
systems in this respect can be colloids where the inter-particle
potential can in principle be tuned to reproduce the SALR potential
\cite{ciach_origin_2013}. However, despite the extensive efforts, no
periodically modulated mesophases have been observed in colloidal
systems. The only mesophases found in these
experiments were equilibrium
clusters\cite{stradner_equilibrium_2004,campbell_dynamical_2005,groenewold_kegel_2001,zhang_polydispersity_2015}
and disordered
gels.\cite{sciorti2004,coniglio_percolation_2004,coniglio_modulated_2011,liu_colloidal_2019}
The failure of producing periodically modulated microphases in
colloids was attributed to the sluggish dynamics of these
systems. The situation in simulations has not been very different.
SALR fluids have been shown to form equilibrium
clusters,\cite{mossa_ground-state_2004,sciortino_one-dimensional_2005,sweatman_2014,sweatman_lue_2016,bollinger_fluids_2016}
and lamellar and columnar microphases have been observed in Monte
Carlo simulations\cite{decandia_2006} and predicted by
density-functional theory.\cite{archer_model_2007,bomont_theoretical_2012}
The complete equilibrium phase diagram of a continuous-space SALR
model, including a double-gyroid phase, was obtained by free-energy
calculations in which each candidate morphology was stabilised by a
modulated external field imposing its
symmetry;\cite{zhuang_prl_2016,zhuang_communication_2017} the
networks did not self-assemble spontaneously. Isotropic pair
potentials producing porous mesophases with a prescribed pore size
have also been obtained by inverse
design.\cite{lindquist_assembly_2016,lindquist_interactions_2017,lindquist_role_2019}
The field has been reviewed
recently.\cite{zhuang_recent_2016,charbonneau_zhang_2022} Molecular
dynamics simulations using simple systems with SALR-type interaction
potentials can thus be used to produce an important insight into the
relevant microscopic-scale formation mechanisms.

In this contribution we review the recent advances in the molecular
dynamics simulations of the microphase separation phenomena in simple
systems using SALR pair potentials and describe periodic mesophases
produced by these simulations. We present a prototype functional form
of SALR pair potential that is shown to be able to produce the self-assembly
of a wide range of periodic mesophases. Besides those
commonly observed in liquid crystal systems and polymers, several new
phases were discovered. It is shown that for this class of pair
potential a first-order microphase separation transition from
the isotropic liquid takes place in agreement with the hypothesis of
Lebowitz and Penrose.  Further, it is shown that some of these phases
are of fluid nature while maintaining their periodicity at the
macroscopic scale, thereby supplanting the spinodal domain predicted
by the van der Waals model.  In this discussion we will particularly
focus on the formation of a novel triply periodic mesophase possessing
the $Fddd$ symmetry. The fact that a system of particles with
spherically symmetric interaction can form a periodic mesophase of
such complexity explicitly demonstrates that the anisometric shape of
the constituent molecules/particles is not a prerequisite for the
self-assembly of periodic microphases.

The role of the simulations using simple particle systems in the
studies of liquids deserves a special remark. The literature dealing
with dense liquids displays a major lacuna in comparing the
theoretical calculations and the appropriate experimental results. In
turning away from intuition and emphasising rigour, a theoretician
relies on the studies of simple systems to maintain the connection
with the real world. The particle simulations of simple systems are
therefore indispensable in bridging the gap between theoretical
concepts and the pertinent experimental data.

 A terminological remark needs to be made regarding the nomenclature
 of the microphases we discuss here. According to the concept of
 ``Copernican crystallography'' introduced by Mermin
 \cite{mermin_copernican_1992} periodic phases should be classified
 according to the symmetry of their reciprocal-space patterns rather
 than to the symmetry invariance of the real-space configurations. It
 will be shown that in terms of the diffraction patterns the lamellar
 and columnar phases discussed here, produced in one-component simple
 particle systems, are indistinguishable from the smectic and columnar
 liquid crystals, respectively. We therefore assume that both
 classifications are equally applicable for describing these
 phases. For the same reason we do not discriminate in this discussion
 between microphases and mesophases.

\section{Simulation}\label{sec:sim}

\subsection{SALR pair potentials}

The original pair potential suggested by Lebowitz and Penrose \cite{lebowitz_rigorous_1966} as a
candidate for producing mesoscopic-scale liquid-gas phase separation
was composed as a superposition of two Kac potentials \cite{Kac}.  Later this
kind of potential was considered in Landau
theory\cite{brazovskii_1975,ciach_universal_2008,ciach_origin_2013,muratov_theory_2002}
and in density-functional
models\cite{wu_dft_2009,archer_model_2007,pini_2015,edelmann_gyroid_2016}
demonstrating the possibility of the formation of periodic microphases
in simple liquids.

\begin{figure}[htbp] %figure 1
\includegraphics[width=1.0\columnwidth] {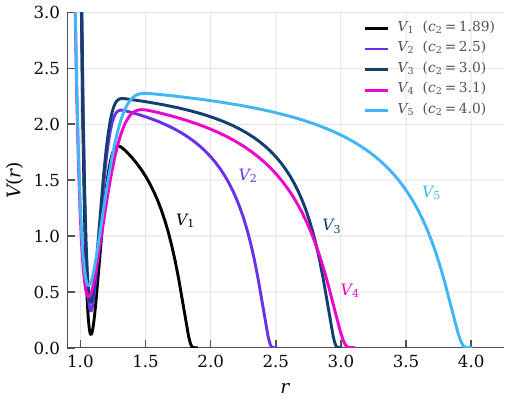}
\caption{ The pair potentials used in the simulation models we
  discuss in this article. The functional form of the potentials is
  described by Eq.~(\ref{potential}). Each potential is defined using the respective
  set of parameters presented in Table \ref{table_pot}. Within each of the
  two families ($V_1$--$V_3$ and $V_4$--$V_5$) the curves differ only in the
  range $c_2$ of the long repulsion, which sets the mesoscopic length scale
  of the resulting morphology.}
\label{fig1}
\end{figure}

We consider here molecular dynamics simulations where
the potential energy for two particles separated
by the distance $r$ is described by the following SALR functional form:

\begin{eqnarray}  \label{potential}
V(r)  = a_1  ( r^{-m} - d) H(r,b_1,c_1) + a_2 H(r,b_2,c_2)\\ \nonumber
H(r,b,c) = \left\{ \begin{array}{ll}
  \exp\left( \frac{b}{r-c} \right) & r < c\\
0 & r \geq c
\end{array}
\right.
\end{eqnarray}

The pair potentials generated by this functional form, which we
discuss in this article, are shown in Fig.~\ref{fig1}, and the
respective sets of parameters are presented in Table~\ref{table_pot}. Like
all SALR potentials they feature a short-range core repulsion and a
short-range attraction, followed by a long-range
repulsion. An important convenience of this functional form, which can
be seen in Eq.~(\ref{potential}), is that it allows the three parts of the
potential to be changed independently of each other
by appropriate variations of the parameters.

\begin{table} [h!]
\begin{tabular}{ccccccccc}
 \hline
\hline \\

& m \space  & $a_1$ \space & $b_1$ \space & $c_1$ \space & $a_2$
\space & $b_2$ \space & $c_2$ & \space $d$ \\
\\
\hline \\
$V_1$ \space \space & 12 \space \space & 265.85 \space \space & 1.5
\space \space & 1.45 \space & 2.5 \space & 0.19 \space &
1.89 \space & 0.8\\
\hline
\hline \\
$V_2$ & 12 \space & 265.85 \space & 1.5 \space & 1.45 \space & 2.5 \space & 0.19 \space &
2.5 \space & 0.8\\
\hline \hline \\
$V_3$ & 12 \space & 265.85 \space & 1.5 \space & 1.45 \space & 2.5 \space & 0.19 \space &
3.0 \space & 0.8\\
\hline \hline \\
$V_4$ & 12 \space & 113 \space & 2.8 \space & 1.75 \space & 2.57 \space & 0.3 \space &
3.1 \space & 1.4\\
\hline
\hline \\
$V_5$ & 12 \space & 113 \space & 2.8 \space & 1.75 \space & 2.57 \space & 0.3 \space &
4.0 \space & 1.4\\
\hline
\hline \\
\end{tabular}
\caption{Values of the parameters for the pair potentials presented
  in Fig.~\ref{fig1}, as published in the original studies: $V_1$, the
  smectic-B crystal\cite{metere_formation_2013}; $V_2$, the \textbf{alm}
  metal-organic framework\cite{metere_mof_2014}; $V_3$, the columnar liquid
  crystal\cite{metere_formation_2015} and the smectic dodecagonal
  quasicrystal\cite{metere_smectic_2016}; $V_4$, the hexatic smectic
  phase\cite{agosta_hexatic_2018} and System~II of the $Fddd$
  study\cite{agosta_self-assembly_2020}; $V_5$, System~I of the $Fddd$
  study\cite{agosta_self-assembly_2020}.}
\label{table_pot}
\end{table}

The described functional form of SALR potential represents an
extension of the pair potential introduced in
Ref.~\onlinecite{dzugutov_icosahedral_1992} and used in an earlier study to
produce a dodecagonal quasicrystal.\cite{dzugutov_qs} That potential
was designed to favour the icosahedral local order by suppressing the
alternative local particle configurations; it was later shown to be a
thermodynamically ideal glass former.\cite{elenius_oppelstrup_dzugutov_2010} The first minimum in the
SALR potentials presented here, Fig.~\ref{fig1}, is designed so
narrow that it also excludes the icosahedral packing of the nearest
neighbours. In this way, the potentials disfavour any conceivable 3D
pattern of the close packing which must typically include 12 nearest
neighbours. It is therefore expected that a local pattern with a
smaller number of neighbours will be formed, while keeping the other
neighbours at a larger distance as defined by the range of the long
repulsion part of the potential. In a simple case, such a structure
can be a stack of flat layers composed of hexagonally packed
particles.

We recall that the goal of this contribution is to review the low
density phases that can be produced in particle simulations using
SALR potentials. In this context, the crucial feature of these pair
potentials is the range of the long-repulsion part which defines the
size of the structural pores in the simulated materials. As one can
see in Fig.~\ref{fig1} and Table~\ref{table_pot}, that part of the SALR
potential can be regulated by the appropriate selection of the potential
parameters without a measurable change in the short-range part. In the following, we will
present the simulation
results\cite{metere_formation_2013,metere_mof_2014,metere_formation_2015,metere_smectic_2016,agosta_hexatic_2018,agosta_self-assembly_2020} demonstrating that a wide variety of
mesoscopic structures can be produced using the described functional
form by only changing the range of the long repulsion of the pair
potential while keeping practically unchanged the short-range part,
including the potential-energy minimum which corresponds to the
position of the nearest neighbours.

\subsection{Structure characterisation}

The structure analysis of the mesophases produced by the molecular
dynamics simulations using the described SALR pair potentials has been
performed by inspecting the reciprocal-space pattern of the density
distribution of the simulated systems. For that purpose, we calculated
the structure factor $S({\bf Q})$ which represents the scattered beam
intensity as measured in the diffraction experiments. To remove
thermally induced fluctuations, the simulated configurations were
subjected to the steepest descent energy minimisation, which mapped them
onto the nearest minimum of its energy landscape.

The static structure factor $ S({\bf Q})$ is defined as
\begin{equation}
  S({\bf Q})= \langle \rho({\bf Q})\rho(-{\bf Q})\rangle
\end{equation}
where $\rho({\bf Q})$ is the ${\bf Q}$-component of the spatial
Fourier-transform of the instantaneous number density distribution of
a system of $N$ particles:
\begin{equation}
\rho({\bf Q},t) = \frac{1}{\sqrt{N}} \sum_{i=1}^{N} \exp [-i{\bf Q r}_i]
\end{equation}
${\bf r_i}$ being the position of particle $i$, and $\langle \
\rangle$ denoting ensemble averaging.\\
A spherically-averaged static structure factor can be calculated from
the radial distribution function $g(r)$ as:
\begin{equation}
  S(Q) = 1 + 4 \pi \rho \int_{0}^{\infty} [g(r)-1] \frac{\sin(Qr)}{Qr} r^2 dr
\end{equation}
As a first step of the structure analysis, $S({\bf Q})$ was calculated
on the reciprocal-space sphere of the radius corresponding to the
position of the first peak of the spherically averaged structure
factor. The pattern of well-defined maxima on the $\bf
Q$-sphere thus defined represents the decomposition of the peak of the spherically
averaged structure factor. This makes it possible to determine the
global symmetry of the simulated particle configuration, and its axis
orientation.

\section{Solid low-density mesophases}\label{sec:solid}

The number of newly discovered low-density porous solid
structures grows rapidly. Description, classification, and design of
these structurally complex materials can arguably be regarded as the
last frontier of crystallography. These systems currently attract a
great deal of research activity because of both their conceptual
interest and the vast scope of their technological
applications. The specific molecular composition of
porous low-density materials is quite complex, which makes it
difficult to analyse the mechanisms of their formation. Therefore it
is of great interest to model the formation of these materials using
simulations of simple systems, specifically as a result of the microphase
separation transition in systems of particles interacting via SALR
pair potentials.

As we have mentioned in the introduction, SALR pair potentials,
besides replacing the macroscopic spinodal coexistence of liquid and
gas bulk phases by a single macroscopically uniform mesophase, can also
break up the macroscopic-scale coexistence between a bulk crystalline
phase and a fluid phase (gas), forming a microphase with a crystalline
morphology. We shall discuss here three kinds of such morphologies
spontaneously formed in MD simulations. The first is a solid phase
with hexagonally close-packed layers where the interlayer distance
greatly exceeds the interparticle distance within the layers. Structures
of this kind are formed in mesogenic systems and are named smectic-B
crystals, or crystal-B phases in the IUPAC nomenclature.\cite{baron_iupac_2001} These are genuine crystals possessing global
3D periodicity where axially stacked layers are composed of elongated
mesogenic particles.\cite{chandrasekhar,LE1} The singular feature of these phases which distinguishes them from common hexagonal close-packed crystalline
structures is that, because of the elongated shape of their
constituent mesogenic particles, the interlayer spacing significantly
exceeds the interparticle distance within a layer. The term
``smectic'' was coined by G. Friedel\cite{friedel_1922} to define layered
mesophases exhibiting fluidity along the layers (now called Smectic
A). Although smectic crystals show a finite elastic response to a
shear applied parallel to the layers, the latter is several orders of
magnitude smaller than their other elastic constants.\cite{CAG,OST}
This large-scale elastic anisotropy is another distinctive anomaly
discriminating Smectic crystals from other crystalline phases. In
this sense smectic-B crystals can be classified as \textit{soft crystals}. 

The elastic anisotropy of real smectic-B phases is caused by the
elongated shape of the constituent molecules, which defines the
interlayer spacing and makes the interlayer cohesion forces much
weaker than the interparticle forces within a layer. In the smectic-B
phases simulated using the SALR potential this feature is provided by
the difference between the strong interparticle cohesion arising from
the steep potential energy well in the short-range part of the
potential and the less steep long-range repulsion, which determines the
interlayer spacing and the interlayer cohesion.

Historically, several highly ordered materials were
misclassified as smectic liquid crystals because of their visual
textures under the microscope; the International Union of Pure and
Applied Chemistry (IUPAC) subsequently clarified that such structures
possessing strict three-dimensional long-range order are in fact soft
crystals, or crystalline mesophases.\cite{baron_iupac_2001,kato_softcrystals_2019}
Outside the strict nomenclature the term also describes soft,
responsive crystalline systems such as metal-organic frameworks
(MOFs).\cite{kato_softcrystals_2019} Another soft crystal induced by a SALR potential in a simple system of
particles, which we discuss below, has the topology of a MOF. Like smectic-B crystals, real MOFs include elongated building units which give rise to large-scale pores. In the simulated MOF these are imitated by the long-range repulsion part of the SALR potential.
Finally, we also discuss here a very peculiar kind of solid smectic mesophase
that was produced in a simulation using a SALR pair potential, a
smectic dodecagonal quasicrystal. While having the same layered
uniaxial structure as the smectic-B crystal, it exhibits 12-fold
rotational symmetry. This structure represents a novel kind of
quasicrystal featuring conceptually interesting anomalies.

\subsection{Smectic-B crystal}\label{sec:cryb}

It was demonstrated that the free energy of a system of particles
using a SALR potential may possess a minimum which corresponds to a
solid lamellar phase with crystalline order within the
layers.\cite{tarzia_pattern_2006,tarzia_lamellar_2007} However, the simulations failed to produce the spontaneous formation of this phase.\cite{decandia_2006}
So far, colloidal smectic phases have only been found in systems of
rodlike colloidal particles.\cite{maeda_2003}
We discuss here a molecular dynamics
simulation using the pair potential $V_1$ described in Sec.~\ref{sec:sim}, which was found to produce the spontaneous formation of a
smectic-B crystal from the uniform fluid
phase.\cite{metere_formation_2013} The simulation was carried out at
the constant number density $\rho=0.55$. We note that this
density is much below the triple-point density for the Lennard-Jones (LJ) system.\cite{hansen_theory_2013} The latter is a useful reference system since, as
we mentioned earlier, the short repulsion part of the pair potential
approximates that of the LJ potential. A first-order transition from the
liquid state into a solid phase upon isochoric cooling was observed.

\begin{figure}[tbp] %figure 3
\includegraphics[width=\columnwidth]{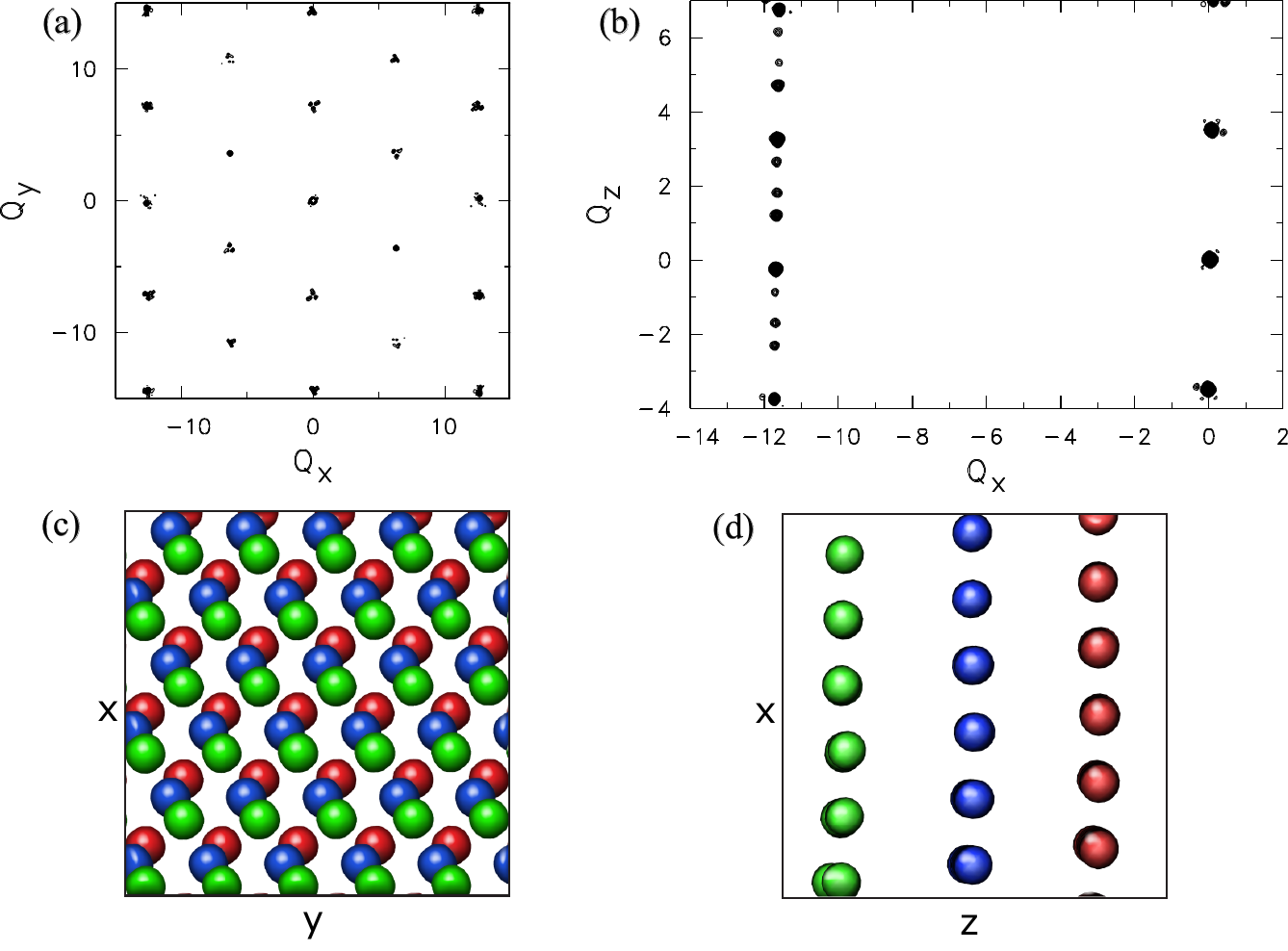}
\caption{The isointensity plots of the structure factor S(Q), in two
orthogonal Q-space planes. (a) Qz = 0; (b) Qy = 0. Qz denotes the
axial dimension, and Qy corresponds to a translational symmetry
vector, orthogonal to the axis. A fragment of the simulated smectic-B
configuration comprising three adjacent layers stacked in ABC-type
sequence. Left: axial view; right: orthogonal view along an in-layer
translational symmetry direction. The layers are distinguished by colour. Adapted with permission from Ref.~\onlinecite{metere_formation_2013}. Copyright 2013 by the American Physical Society.}
\label{smecticB}
\end{figure}

The structure analysis of the solid phase arising upon cooling has been
performed by inspecting the Fourier-space pattern of its density
distribution. For that purpose, the structure factor $S({\bf Q})$
representing the diffraction intensity was first calculated on the
$Q$-space sphere corresponding to the first peak of the spherically
averaged $S({\bf Q})$.  Well-defined diffraction maxima forming a
regular pattern were observed, which made it possible to determine the
global symmetry of the configuration; a single hexagonal axis was
detected. The axis orientation having been found, $S({\bf Q})$ was
calculated within two characteristic $Q$-space planes: $Q_z=0$ and
$Q_y=0$, $Q_z$ being the axis coordinate and $Q_y$ coordinate
corresponding to a translational symmetry vector orthogonal to the
axis. The two diffraction patterns are shown in Fig.~\ref{smecticB}.

This structure analysis compels us to conclude that the low-temperature
phase produced by the described phase transition is a true 3D crystal
which is indicated by the observable sharpness of the diffraction
maxima. Moreover, the two diffraction patterns exhibit structural
features characteristic of the experimentally produced smectic-B
phase:\cite{LE2,LE1} the configuration is a uniaxial crystal comprised of stacked
layers with dense hexagonal packing of particles in each layer. Based
on the diffraction results shown in Fig.~\ref{smecticB}, we can also estimate the ratio of the interlayer distance to the nearest-neighbor separation
within a layer as 1.8. These distances apparently correspond,
respectively, to the short-repulsion and long-repulsion parts of the
pair potential. The ratio of these distances can be associated with
the aspect ratio of the mesogenic particles forming the experimentally
produced smectic-B crystal, which defines the interlayer distance.

The diffraction data presented in Fig.~\ref{smecticB} also provide comprehensive
information concerning the interlayer correlations in the crystal. The
hexagonal arrangement of sharp diffraction peaks in the $Q_z = 0$ plane
implies the existence of global positional interlayer
correlations. The information about the type of positional
correlations of the particles of adjacent layers can be obtained by
inspecting the pattern of diffraction intensity in the axial plane,
Fig.~\ref{smecticB}. The hexagonal close-packed layers may be stacked with three
possible ordered arrangements: AAA\ldots, ABA\ldots, or ABCA\ldots, where A, B,
and C denote the relative position of the layers. A random array
of ABC-type planes is also possible. These types of layer packing, or
their mixtures have been experimentally observed.\cite{LE2}  The diffraction
intensity profile along the axial coordinate in Fig.~\ref{smecticB} demonstrates
four distinct auxiliary peaks, interposed between the main peaks which
represent the general layers' periodicity. This pattern can be
interpreted as representing the predominantly ABCA-type of order in
layers' packing with possible defects.\cite{LE2} An example of this type of
arrangement of adjacent layers discerned by the real-space inspection
of the simulated smectic-B configuration is presented in Fig.~\ref{smecticB}.
We note that the smectic-B structures formed by systems of spherical
particles using SALR potentials are stacks of identical, hexagonally
ordered layers and are therefore, in terms of the diffraction pattern
and disregarding the chemical composition, identical to the smectic-B
crystals of liquid crystals,\cite{LE1,LE2} unlike the lamellar phases of
block copolymers, which are alternating stacks of chemically distinct
layers that are liquid within the layer.\cite{bates_network_2005}

\subsection{Formation of a new archetypal Metal-Organic Framework from
  a simple monatomic liquid}\label{sec:mof}

Metal--organic frameworks
(MOFs),\cite{tranchemontagne_2009,wang_yaghi_2008,banerjee_2008,okeeffe_2009}
a class of porous materials, feature tunable structures and diverse
functionalities for applications in gas separation, storage and
catalysis.\cite{okeeffe_yaghi_2012,wilmer_2012} A key factor
contributing to their performance lies in structural design using
novel topologies.  Reticular synthesis is a crystal engineering
approach whereby pre-designed stable building blocks\cite{tranchemontagne_2009} are assembled
into a preliminarily conceived periodic network. The development of
this approach resulted in the current explosive growth of the number
of newly created MOFs. The design of new MOFs crucially depends on
their modelling and systematic structural characterisation.

The general purpose of mathematical modelling of a phenomenon consists
in the reduction of its observed complexity, which makes it possible
to discern its essential aspects.  In this way, the investigated
phenomenon can be understood by putting it in a general context and
providing its unifying description within an appropriate
classification.

The advent of reticular chemistry was prompted by a principally new
approach to the description of extended low-density periodic
structures. The latter are considered in terms of topology of infinite
periodic graphs (nets) which represent the networks of bonds
connecting the elementary structural blocks. Using the structure
description in terms of the topology of bonds, hypothetical MOFs can
be redesigned from the existing structures by recombining the building
blocks according to the conceived net topology.\cite{okeeffe_yaghi_2012,wilmer_2012}

\begin{figure}[htbp]
\includegraphics[width=0.72\columnwidth]{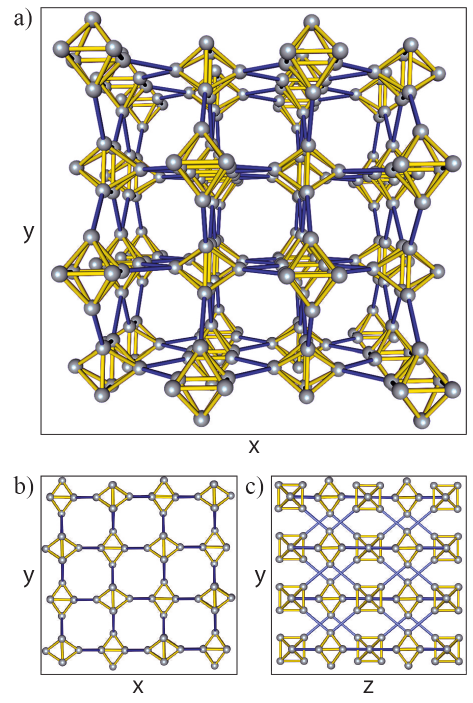}
\caption{A fragment of the simulated crystal configuration produced
  using the same bonding method as used for the unit cell plot in
  Fig.~\ref{mof_unit}, using the same respective bond colours. (a) Perspective
  projection demonstrating how the octahedra are interconnected to
  form the pores. (b) and (c) Orthogonal projections. Reproduced from Ref.~\onlinecite{metere_mof_2014}, with the permission of AIP Publishing.}
\label{mof_frag}
\end{figure}

\begin{figure}[htbp]
\includegraphics[width=0.55\columnwidth]{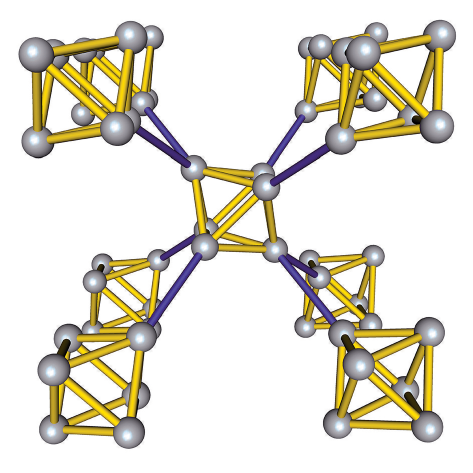}
\caption{The unit cell of the simulated crystal configuration. The
  yellow and purple bonds correspond to the distance ranges
  highlighted by similar colours in Fig.~\ref{mof_frag}. Reproduced from Ref.~\onlinecite{metere_mof_2014}, with the permission of AIP Publishing.}
\label{mof_unit}
\end{figure}

\begin{table}
\begin{tabular}{cccc}
\hline
\hline
Space group & $a=b$ & $c$ & $\alpha=\beta=\gamma$\\
\hline
\hline
\space $P4_{2}/mnm$ \space & 3.600 \space & 3.200 \space & \space 90$^\circ$
\space \\
\hline \\
\end{tabular}

\begin{tabular}{cccc}
\hline
\hline
Vertex & x/a & y/b & z/c\\
\hline
\hline
\space $V_1$ \space & \space 0.39342 \space & \space 0.39342 \space & \space 0.33077 \space \\
\space $V_2$ \space & \space 0.14994 \space & \space 0.14994 \space & \space 0       \space \\
\hline \\
\end{tabular}
\caption{Reconstructed unit cell parameters and list of the vertices in the
asymmetric unit. Reproduced from Ref.~\onlinecite{metere_mof_2014}, with the permission of AIP Publishing.}
\label{table2}
\end{table}

A basic question arising in the evaluation of a hypothetical MOF topology
is whether the atomic configuration designed in that way represents a
free-energy minimum. To answer this question, the use of atomistic
simulation is indispensable. This raises another question of
conceptual interest: how simple could a physical model, reproducing
a MOF topology, be?  The molecular-dynamics simulation which
we describe here\cite{metere_mof_2014} demonstrates that a single-component system of
particles using a SALR pair potential can form a thermodynamically
stable crystalline structure with a net topology class observed in
MOFs.

The simulation was carried out using potential $V_2$ at constant
number density $\rho = 0.3$, in reduced units. By cooling the liquid
below $T = 0.65$, a crystalline phase was produced. Its real-space
configuration, and the unit cell, shown in Figs.~\ref{mof_frag} and
\ref{mof_unit}, respectively,
represent a typical low-density porous MOF structure.  The unit cell
is tetragonal, its parameters and the positions of the unit vertices,
scaled by the size of the unit cell, are reported in Table~\ref{table2}.
Connecting the nearest neighbours with bonds of the length $r_1=1.08$
forms a regular octahedron. These bonds correspond to the minimum
position of the potential. Bonds of the length $r_2 \approx 1.63$
connect the central octahedron with its replicas at the corners. Their
self-repetition along the axes generates a one-dimensional channel
system. The pore diameter estimated as $ \approx 2.5$, is consistent
with the range of the long-repulsion part of the potential.

This is a new archetypal MOF structure. Its novel net topology has
been included in the RCSR database\cite{rcsr} and denoted by ``{\bf alm}''.

\subsection{Smectic dodecagonal quasicrystal}\label{sec:qc}

Quasiperiodic order in mesoscopic soft-matter systems is a new
research frontier of rapidly growing interest \cite{DUB}.  First of
all, the interest in the studies of mesoscopic-scale soft-matter
quasicrystals is motivated by the prospective application of
mesoscopic quasicrystals in photonics \cite{VARD}.  Following the
first discovery of a dodecagonal quasicrystal in a micelle-forming
system,\cite{ZENG} similar structures were produced in several
soft-matter systems\cite{FISCH,CHAN,TAL,DO2} and in
simulations.\cite{GLOTZ} Besides tetrahedrally close-packed structures
characteristic of intermetallic compounds, other types of
quasiperiodic order were observed.\cite{BARK1,BARK2}

\begin{figure}[tbp]
\centering
\includegraphics[width=0.8\columnwidth]{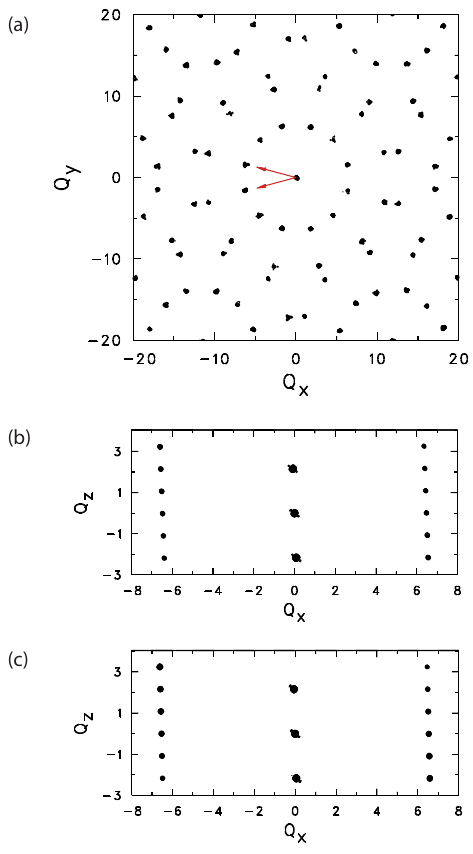}
\caption{Isointensity contour plots of the structure factor $S({\bf Q})$. (a) In
the $Q$-plane perpendicular to the axis. (b) and (c) represent respectively
two $Q$-planes defined by the axis and the two vectors, $30^{\circ}$ apart, shown
in panel (a). Reproduced from Ref.~\onlinecite{metere_smectic_2016} with permission from the Royal Society of Chemistry.}
\label{qc_sq}
\end{figure}

The simulation we discuss here\cite{metere_smectic_2016} used the
potential $V_3$ at the number density
$\rho=0.32$. Upon isochoric cooling from the equilibrium liquid state
the system underwent a first-order transition to a solid phase below
$T=1.1$. To detect the symmetry and orientation of the produced
structure, its $S({\bf Q})$ was analysed in the same way as for the
smectic-B crystal; the resulting pattern in the ${\bf Q}$ plane perpendicular to the axis is shown in Fig.~\ref{qc_sq}a.
The pattern of $S({\bf Q})$ maxima in that plane is
apparently 12-fold. The 12-fold rotational invariance of the simulated
configuration is also apparent from the comparison of the $S({\bf
  Q})$- patterns in two planes defined by the axis and two
translational symmetry vectors, $30^{\circ}$ apart, shown in
Figs.~\ref{qc_sq}b and \ref{qc_sq}c.

\begin{figure*}[tbp]
\centering
\includegraphics[width=0.42\textwidth,viewport=0 238 245 468,clip]{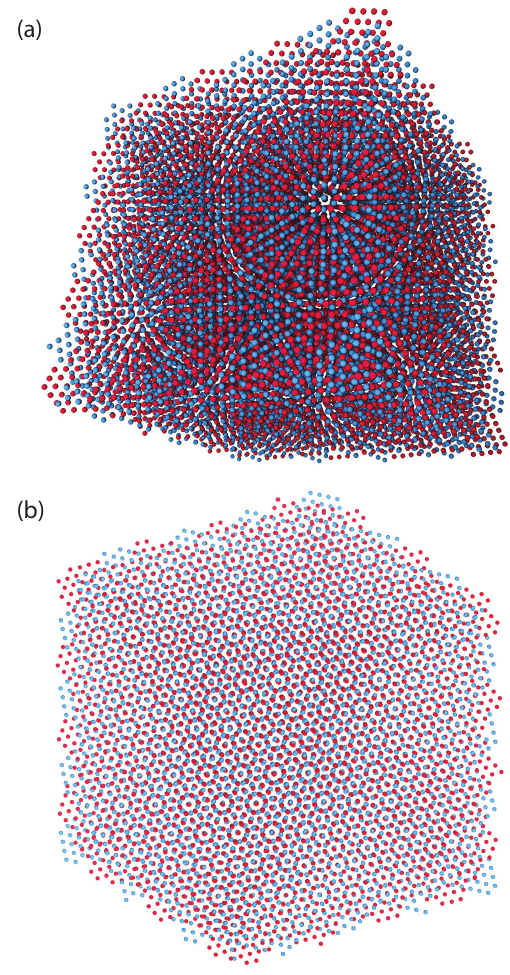}\hfill
\includegraphics[width=0.42\textwidth,viewport=0 0 245 238,clip]{fig4.pdf}
\caption{An axial view of the configuration. (a) Perspective
  projection; (b) orthogonal projection. Adjacent A and B layers are
  discriminated by colour. Reproduced from Ref.~\onlinecite{metere_smectic_2016} with permission from the Royal Society of Chemistry.}
\label{quasicrystal_real}
\end{figure*}

\begin{figure}[htbp] %figure 3
\includegraphics[width=0.8\columnwidth]{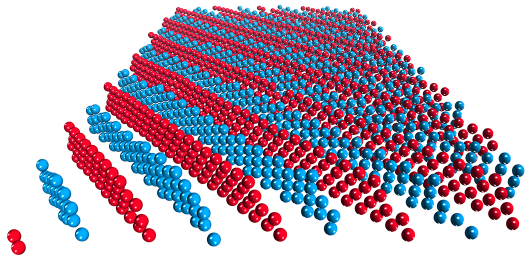}
\caption{ A slice of the configuration cut parallel to the axis. A and B layers
are marked by red and blue colours. Reproduced from Ref.~\onlinecite{metere_smectic_2016} with permission from the Royal Society of Chemistry.}
\label{slice}
\end{figure}

\begin{figure}[htbp] %figure 3
\includegraphics[width=0.62\columnwidth]{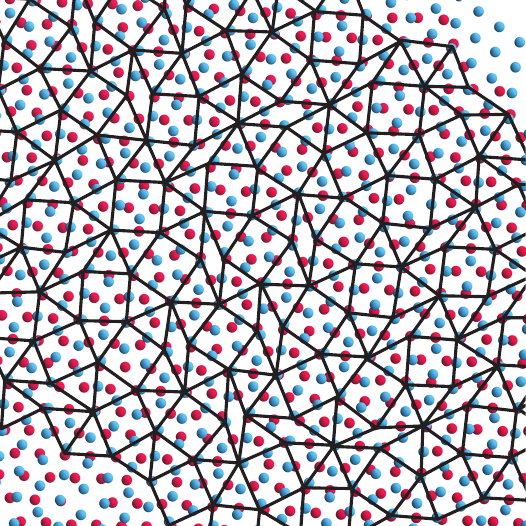}
\caption{ Tiling of the axial projection of two adjacent particle layers,
produced by connecting centres of adjacent 12-particle rings. Reproduced from Ref.~\onlinecite{metere_smectic_2016} with permission from the Royal Society of Chemistry.}
\label{tiling}
\end{figure}

\begin{figure}[htbp] %figure 5
\includegraphics[width=0.6\columnwidth]{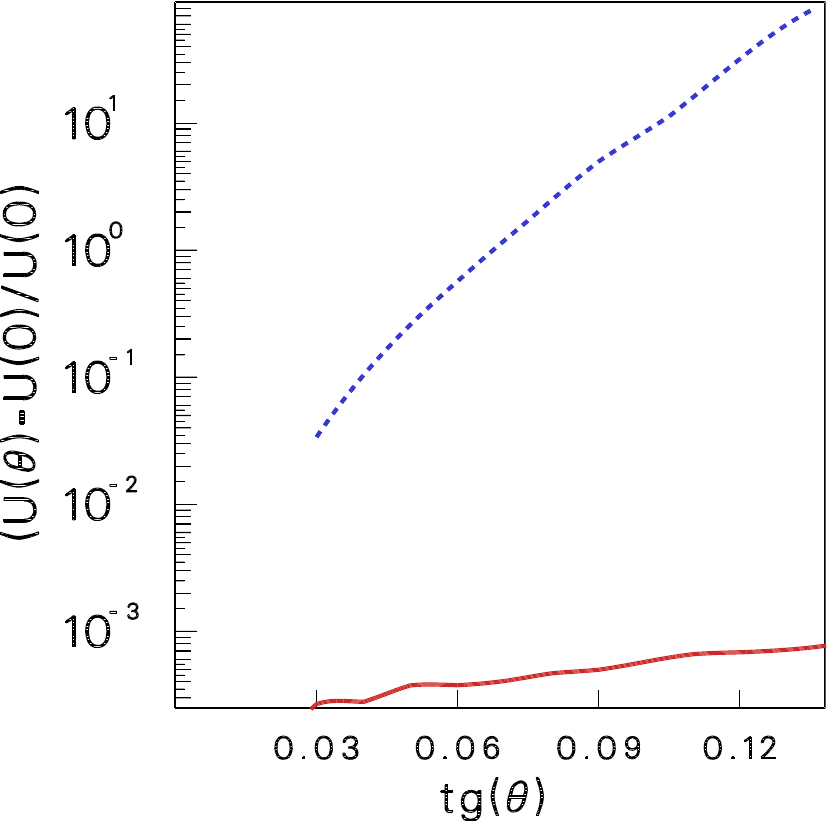}
\caption{ Variation of the system's potential energy $U$ as a response
  to shear strain applied in the $XZ$ plane (solid line), and in the
  $XY$ plane (dashed line). $Z$ is the stacking axis direction, $X$
  corresponds to a symmetry direction in the layer plane. Reproduced from Ref.~\onlinecite{metere_smectic_2016} with permission from the Royal Society of Chemistry.}
\label{qc_shear}
\end{figure}

Besides, the latter two diffraction patterns look like those observed
in typical smectic-B crystals \cite{metere_formation_2013,LE1}. They represent a
structure composed of flat close-packed particle layers axially
stacked in $ABA$ order \cite{LE2}. The diffraction results in
Fig.~\ref{qc_sq}
also make it possible to estimate the interlayer spacing and the
interparticle distance within a layer as $2.95$ and $1.1$; their ratio
can be compared with the aspect ratio of constituent particles in
commonly occurring smectic phases. We note that this value is
consistent with the ratio of the positions of two minima in the pair
potential, see Fig.~\ref{fig1}.
Thus, the configuration is structurally similar to the smectic-B
crystal, except that, in contrast to hexagonal symmetry of the latter,
it exhibits 12-fold symmetry.

These diffraction results can be interpreted by assuming that the
system's apparent 12-fold axial symmetry is induced by
$30^{\circ}$ rotation of the $A$ and $B$ subsystems of layers
with respect to each other, in the layer plane. This conclusion
is supported by the visual inspection of the real-space images of
the simulated structure presented in Fig.~\ref{quasicrystal_real}.
 The view from the
axial direction shows a pattern of 12-particle rings
characteristic of dodecagonal structures \cite{GA,dzugutov_qs}, and its
particle density lines that can be detected by looking at the
grazing angle exhibit $30^{\circ}$ rotational invariance. When
viewed along the layer plane, Fig.~\ref{slice},
 the configuration looks
like a typical smectic-B crystal with $ABA...$ layer stacking;
its only distinction from the latter in that the adjacent $A$ and
$B$ layers are rotated with respect to each other, in the layer
plane, by $30^{\circ}$.

In Fig.~\ref{tiling} we present tiling of the axial projection of two adjacent
layers of the simulated quasicrystal. The tiling is produced by
connecting the centres of adjacent apparent 12-particle rings which
the pattern is composed of. Besides equilateral triangles and squares
typical for periodic tilings, the present tiling includes the
$30^{\circ}$ rhombus, an element specific for a 12-fold tiling pattern
\cite{GA,dzugutov_qs}.

In crystallographic classification, the symmetry of this quasicrystal
is $P12_6/mmc$. The quasicrystal is composed of stacked layers of
$6^3$ nets that each have planar $6mm$ symmetry.  The stacking along
the axis with a $\pi/6$ relative rotation of the adjacent layers
preserve the mirror operations and subsequent layers are related by a
$12_6$ screw operation and a diagonal glide reflection with a $1/2c$
component. Quasicrystallographic aspects of these operations were
discussed earlier.\cite{MER1,MER2}

Thus, we found a new type of quasicrystalline phase, which can possibly
be classified as a {\it dodecagonal smectic-B} phase. This
quasicrystal is structurally distinct from the earlier reported
quasiperiodic structures produced using spherically symmetric
interparticle potentials \cite{dzugutov_qs,LO,EN,DO}.

As recalled in Sec.~\ref{sec:solid}, the shear elastic constant of smectic-B phases
related to the gliding of the layers is several orders of magnitude
smaller than the other elastic constants. This
anomaly can be seen in the variation of the system's potential energy
as a function of shear strain applied in two planes, $XZ$ and $XY$,
$Z$ being the stacking axis direction, $X$ corresponding to a symmetry
direction in the layer. The degree of elastic anisotropy in
Fig.~\ref{qc_shear} is consistent with that observed in smectic-B phases
\cite{CAG,OST}. It shows that this quasicrystal is a mechanically
stable configuration representing a potential-energy landscape
minimum.

A unique and conceptually significant anomaly of this quasicrystal
is the absence of phason disorder, i.e. of the local atomic
displacements violating the tiling pattern which are characteristic of
other quasicrystals.\cite{DZ3} This anomaly appears to be generic for this type of quasicrystal due
to its physical nature, and the layer stacking order. That the layers
are perfectly identical can be concluded by inspecting the general
view of the configuration from the axial direction which is presented
in Fig.~\ref{quasicrystal_real}.
Due to the strong interaction between particles within a layer induced
by the first minimum in the potential, Fig.~\ref{fig1},
the hexagonal layer structure is solid, making the energy cost for
intralayer dynamics forbiddingly high. Thus, this quasicrystal
structure appears to correspond to a single energy minimum.

The conceptual significance of this peculiarity is related to the
general mechanism of growth and stabilisation of quasiperiodic order.
Since the long-range quasiperiodic order cannot be produced by the
repetition of a unit cell as in conventional crystals\cite{levine_steinhardt_1986}, the growth of
quasiperiodic long-range order appears to require the atoms adhering
to the growth front to obtain nonlocal structural information, which
is physically implausible. This problem has attracted much attention
ever since the first discovery of a quasicrystal in 1984 \cite{shechtman}, and
several theoretical growth models have been proposed. The debates
revolve around the relative roles of energy and entropy.\cite{henley_1991,joseph_elser_1997} The
energetic mechanism postulates that the interactions among the tiles
are such as to favour unique tiling rules. This idea is apparently
inconsistent with Penrose's theorem,\cite{penrose_1989} which states that Penrose
tiling cannot be grown by local rules.\cite{onoda_1988} The entropic mechanism, on the
other hand, assumes that instead the long range quasiperiodic order
arises from entropy maximisation in the ensemble of tilings
randomly changing by uncorrelated phason flips\cite{joseph_elser_1997}.  The
phason-assisted relaxation is thus a necessary mechanism of growth and
stabilisation of ideal quasicrystalline order in real materials. Direct
observation of the growth of a decagonal quasicrystal by in situ
electron microscopy showed that frequent error-and-repair
phason-assisted events are essential for reaching the final
quasiperiodic order\cite{nagao_2015}.

The results described here appear to present a third mechanism of quasicrystal growth and stability, alternative to both: the quasiperiodic order is encoded in the relative rotation of two sublattices of perfectly periodic solid layers, so that no phason degrees of freedom are available and no repair is required.

\begin{figure*}[htbp] %figure 3
\includegraphics[width=0.8\textwidth]{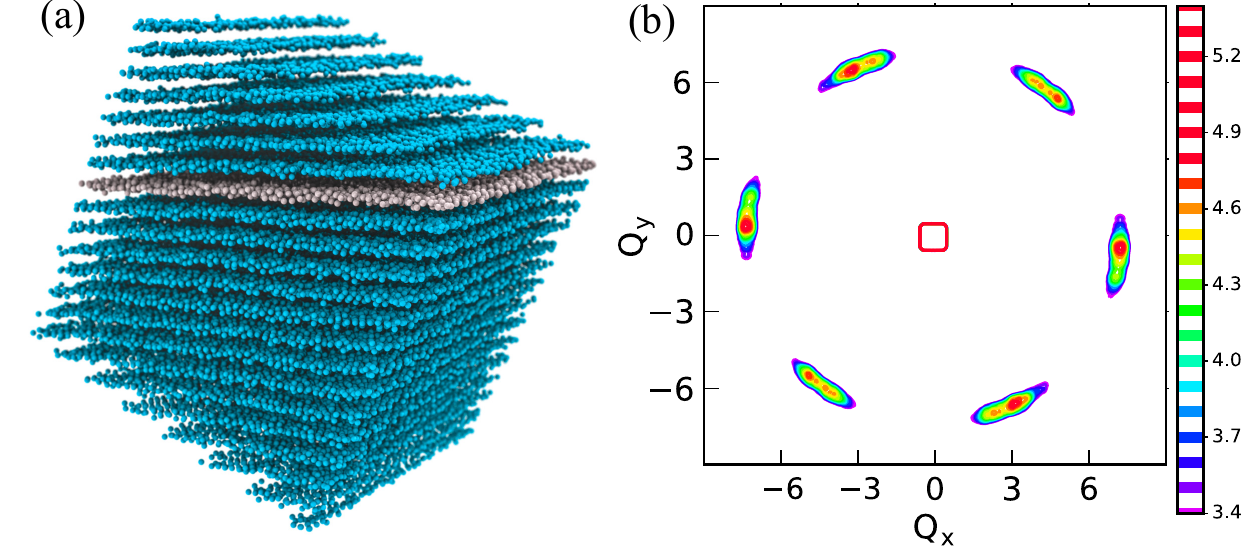}
\caption{ (a) Hexatic-smectic phase. (b) The classical diffraction pattern of a layer of the hexatic-smectic phase. It resembles perfectly the diffraction pattern of 2D hexatic phases.\cite{jin_nature_1996} It is characterised by a six-fold symmetry, indicating the long-range BOO correlation. The peaks are blurred due to the decay in positional order (in a crystal they would be delta functions). Adapted with permission from Ref.~\onlinecite{agosta_hexatic_2018}. Copyright 2018 by the American Physical Society.}
\label{hexatic}
\end{figure*}

\begin{figure}[htbp]
\includegraphics[width=0.78\columnwidth]{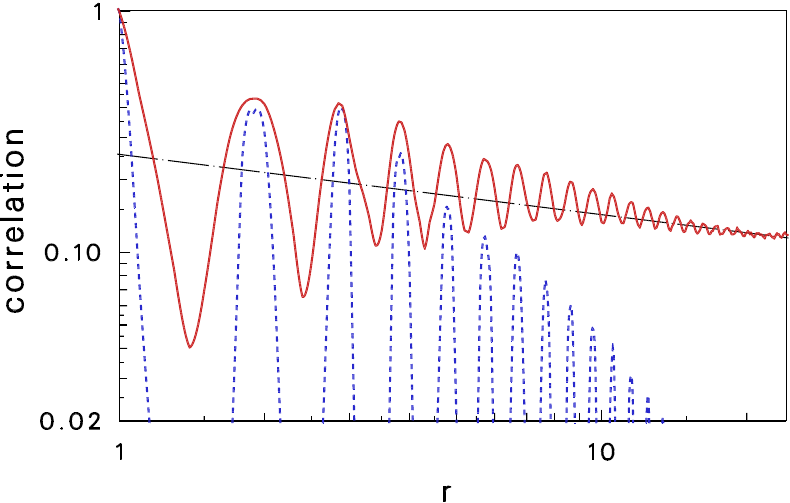}
\caption{Solid line: BOO correlation function $g_6(r)$; dashed line:
  $g(r)-1$, both at $T = 1$; dash-dotted line: an asymptotic
  fit by power law. Reprinted with permission from Ref.~\onlinecite{agosta_hexatic_2018}. Copyright 2018 by the American Physical Society.}
\label{algebraic}
\end{figure}

Two possible ways of materialisation of this structure in real
physical systems can be conjectured. First, the morphological
similarity of the simulated smectic quasicrystal and the smectic-B
crystal suggests that some mesogens forming that phase may also form
the structure we report here.

Second, the pair potential used in the present model shows similarity
to that predicted for colloidal systems by
Deryagin-Landau-Verwey-Overbeek theory \cite{VO,IS} (amended with hard
core repulsion or steric repulsion at close to contact). This suggests
a possibility that smectic-like layered structures exhibiting
quasiperiodic symmetry can be produced in colloidal systems of
spherically shaped particles, with appropriate tuning of the effective
potential.

\section{The hexatic smectic phase}

The hexatic phase is an intermediate phase between the solid and
liquid phases. It was discovered and described by Kosterlitz,
Thouless, Halperin, Nelson and Young
(KTHNY)\cite{kosterlitz_notitle_1972,halperin_notitle_1978,young_notitle_1979,nelson_halperin_1979}
for two dimensions. The hexatic phase is characterised by
quasi-long-range (algebraically decaying) bond-orientational order
(BOO), whereas its short-range positional correlations decay
exponentially, which is typical of standard fluids. Experimentally,
the hexatic phase has been observed in 2D systems of spherical
colloidal particles,\cite{marcus_notitle_1996} but it had never been
found in a layer of a 3D smectic lamellar phase. Three-dimensional
hexatic-B liquid
crystals\cite{jin_nature_1996,Varta1,Varta2,jw_goodby_notitle_1989}
exhibit a true long-range BOO because of the coupling between the
layers.

We discuss here the first reported occurrence of hexatic order
in a 3D smectic phase.\cite{agosta_hexatic_2018} The latter was
obtained in the MD simulation of a system of particles interacting via
the SALR potential $V_4$, see Sec.~\ref{sec:sim}. The system's phase
behaviour was investigated at a constant number density
$\rho=0.41$. A first-order transition to a smectic phase, shown in
Fig.~\ref{hexatic}, occurred under isochoric cooling from the isotropic
liquid at the temperature $T=1.15$. We note that the range of the
long-repulsion part of the potential matches the interplanar distance
of the layers. Fig.~\ref{hexatic} presents the diffraction pattern for
the highlighted layer of the smectic phase. The six symmetric peaks
demonstrate the long-range bond-orientational order, while the circular
spread of the diffraction peaks at the characteristic inter-particle
length scale is the result of the liquid-like translational
diffusion. The identifying feature of the hexatic phase according to
the KTHNY theory is the algebraic decay of its BOO, compared with the
exponential decay of the positional correlations. This behaviour of
the simulated system is demonstrated in Fig.~\ref{algebraic}: the
calculated BOO correlation function asymptotically decays as a power
law, in agreement with the prediction of the KTHNY theory for the 2D
hexatic phase, whereas the radial distribution function $g(r)$ decays
exponentially. Testing these effects experimentally is particularly
challenging due to the impossibility of analysing a single layer
embedded in a 3D structure.

 Being the first KTHNY-predicted 2D hexatic phase found in a 3D system,
 this result expands the scope of application of the theory, including to colloidal
 systems.\cite{n_a_garcia_notitle_2017,zac2} The relevance of this
 result goes beyond simple liquids. A recent study of noble gases and
 aluminium confined between graphene sheets\cite{wang_kosterlitz_2025}
 found that, while a confined monolayer melts through the two-step
 KTHNY scenario, multilayer solids pass continuously into an
 intermediate \emph{layered hexatic} phase, persisting up to at least
 twelve layers, before melting discontinuously into the isotropic
 liquid. The crossover from two-dimensional to three-dimensional
 topological defects invoked there is the same mechanism that the
 present SALR system exposes in its simplest form, with the interlayer
 coupling set by the long-repulsion range of the pair potential.

\section{Periodic fluid microphases}

\subsection{Columnar mesophase}

The experimentally observed columnar liquid crystals represent
close packing of parallel columns composed of axially stacked
molecular units. In the plane perpendicular to the column axis, the
column packing forms a regular pattern with two-dimensional
periodicity, whereas the remaining continuous translational symmetry
dimension is directed along the column axis. The columns are commonly
formed by disk-like molecules \cite{BOCK} or wedge-shaped dendrons
\cite{ZENG}. The close packing of columns fixes the position of a
molecule in the plane perpendicular to the column axis, but the
molecules' stacking along the axis is irregular, and the axial
position of a molecule is not defined with respect to its neighbors in
adjacent columns, which gives rise to the continuous translational
symmetry in the axial dimension.

Particle simulations have proved to be an indispensable tool for
understanding the relationship between the phase behaviour and the
molecular-level properties of liquid crystals \cite{care,bates,zannoni}. These simulations have so far followed the phenomenological
paradigm that dominated the science of liquid crystals for decades,
whereby the structural anisotropy of liquid crystals was assumed to be
determined by the anisometric shape of the molecules. Accordingly, the
shapes of the constituent particles in the models of liquid crystals
have been designed to imitate the shapes of the molecules in the
respective mesogens: the smectic phases have been simulated using
rod-like particles \cite{bolhuis,Allen, sarman, sarman2}, and models of
columnar liquid crystals \cite{zannoni, MARTIN, bates2, ANDRI, veerman}
commonly used flat discotic particles or oblate ellipsoids.

That approach is based on the conjecture that the formation
of structurally anisotropic liquid crystals is driven by the
entropic component of the free energy \cite{LUB}. The origin of this
line of thinking can be traced to the seminal work of Onsager
\cite{onsager_notitle_1949} which linked the anisotropy of the equilibrium structure in
a liquid of rods to the entropy of packing.  The conjecture of the
entropic mechanism of liquid-crystal formation has been further
strengthened by simulations using hard anisometric
particles\cite{Allen,veerman,barmes_2003,mederos_2014} stressing the
role of the geometry of excluded volume. Columnar and lamellar
microphases have been observed in SALR colloids as well, in Monte Carlo
simulations of particles with screened electrostatic
repulsion,\cite{decandia_2006} and in density-functional
calculations;\cite{archer_model_2007} the distinctive feature of the
columnar phase discussed below is the liquid structure and the
diffusivity of its columns.

A question of general conceptual interest for the statistical
mechanics of condensed matter is whether the anisometry of the
mesogenic molecules is really a prerequisite for producing structural
anisotropy in liquid phases, as it is conjectured by the entropic
paradigm, or the entropic effects on the structure formation due to
the particle geometry can be compensated by an appropriately designed
spherically symmetric interaction potential. This question is confined
to the positional structural ordering; the nematic orientational ordering
in liquids of anisometric particles is obviously beyond the scope of
present discussion.

This question has been addressed in a molecular-dynamics simulation\cite{metere_formation_2015}
which exploited a one-component system of particles
interacting via the SALR pair potential $V_3$. The system was
equilibrated in the uniform liquid state at the number density
$\rho=0.3$, which was followed by an isochoric cooling. A phase
transition was detected when the system was cooled below $T=1.1$. We
recall that the simulation using the same potential $V_3$ at the
density $\rho=0.32$, discussed above, produced the dodecagonal smectic
quasicrystal.

\begin{figure}[htbp]
\includegraphics[width=0.46\columnwidth]{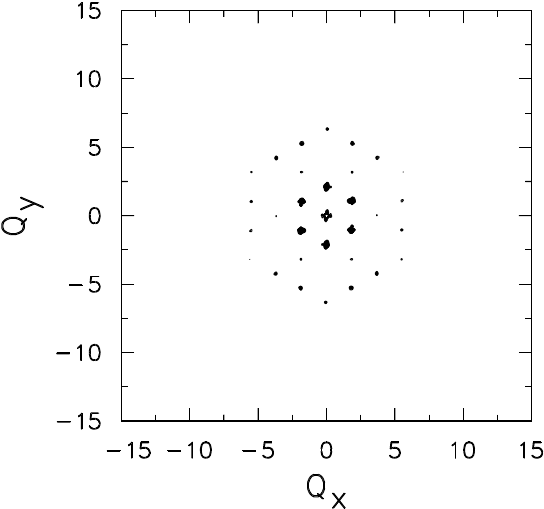}\hfill
\includegraphics[width=0.52\columnwidth]{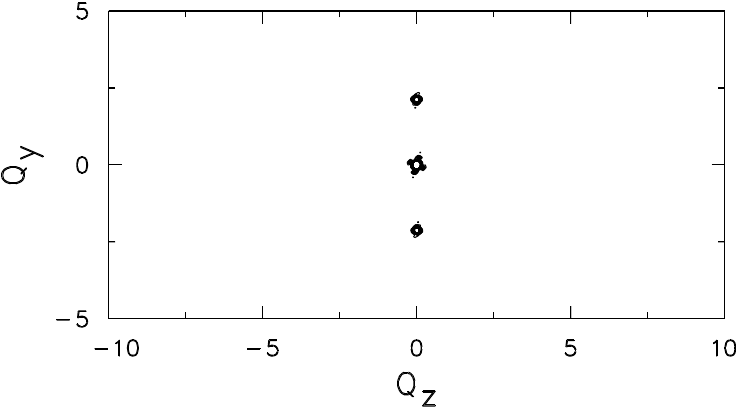}
\caption{ The isointensity plots of the structure factor S(Q)
  calculated for a configuration in two orthogonal reciprocal-space
  planes. Left and right, respectively: $Q_z = 0$ and $Q_x = 0$. Qz
  denotes the axial dimension, and Qy corresponds to one of the
  translational symmetry vectors orthogonal to the axis. Reproduced from Ref.~\onlinecite{metere_formation_2015} with permission from the Royal Society of Chemistry.}
\label{col_sq}
\end{figure}

\begin{figure}[htbp]
\includegraphics[width=0.85\columnwidth]{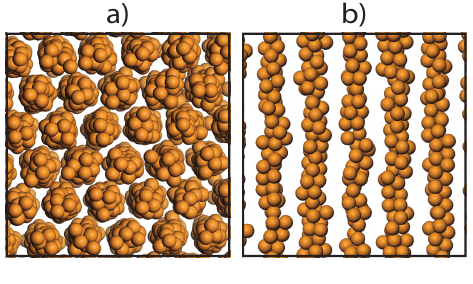}
\caption{ The real-space images of the column phase configuration. (a)
  View from the axial direction. (b) A one-column layer of the
  structure, cut parallel to the column axis, as viewed perpendicular
  to the axis. The particle diameter in the plot is assumed to be 1, in
  the simulation reduced units, which corresponds to the separation
  distance of the particles at the hard-core contact, see the pair
  potential in Fig.~\ref{fig1}. Reproduced from Ref.~\onlinecite{metere_formation_2015} with permission from the Royal Society of Chemistry.}
\label{col_real}
\end{figure}

\begin{figure}[htbp]
\includegraphics[width=0.7\columnwidth]{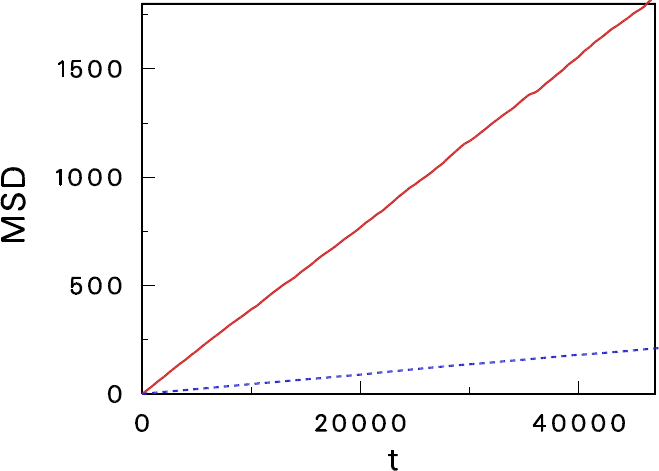}
\caption{ Mean-square particle displacement (MSD) in columnar phase as
  a function of time at T = 1. Solid line corresponds to the diffusion
  along the column axis. Dashed line: MSD averaged over directions
  perpendicular to the axis. Reproduced from Ref.~\onlinecite{metere_formation_2015} with permission from the Royal Society of Chemistry.}
\label{col_msd}
\end{figure}

The structure of the new phase was analysed in reciprocal space in
the same way as for the smectic-B crystal (Sec.~\ref{sec:cryb}); the symmetry
axis having been found, $S({\bf Q})$ was calculated in the planes
$Q_z=0$ and $Q_x=0$, $Q_z$ being the axis coordinate and $Q_y$ the
coordinate along a translational symmetry vector orthogonal to the
axis. The
results in Fig.~\ref{col_sq} show that this is a uniaxial structure
with hexagonal symmetry in the plane perpendicular to the axis,
whereas no structure has been found in the axial dimension. It
consists of columns packed in a six-fold pattern. The
intercolumn separation as inferred from the diffraction pattern is
consistent with the long-range repulsion range of the pair potential,
Fig.~\ref{fig1}.

A remarkable structural aspect that can be
observed in the images of Fig.~\ref{col_real} is that the configuration of a column
can be described as three-dimensional liquid-like dense random
particle packing. This column structure is distinctly different
from that ubiquitously occurring in discotic liquid crystals where the
columns are composed of axially stacked disc-like particles. The
discotic columns are of one-particle width, and they are densely
packed laterally, forming a periodic pattern in the plane
perpendicular to the axis.

The characteristically liquid structure of the columns of this phase
suggests that this phase should also be expected to sustain a kind of
liquid-like dynamics. Indeed, a considerable rate of intracolumn
liquid diffusion has been observed, see Fig.~\ref{col_msd}.

Driven by this
diffusion mechanism, a particle is expected to be able to diffuse to
unlimited distance within a column along the axis, whereas its
transition perpendicular to the axis is limited by the column
diameter. A particle can, nevertheless, leave the column by hopping to
an adjacent one; these hopping events, however, are expected to be
rare relative to the particle intra-column diffusive movements because
of the large energy cost of crossing the potential-energy maximum
separating the columns, see Fig.~\ref{fig1}.

A conceptually significant implication of the present
result is that it directly contradicts the entropic paradigm that
dominated the science of liquid crystals for decades, assuming that the
formation of long-range order in liquids is driven by the entropy
arising from the packing of anisometric constituent
particles.\cite{LUB} This result demonstrates that an anisotropic pattern with two-dimensional
global periodicity can emerge in a liquid composed of a single sort of
particles interacting via a spherically-symmetric potential. Moreover,
this liquid phase has also demonstrated a strongly anisotropic
self-diffusion. This result thus makes it possible to conclude that
emergence of structural and dynamic anisotropy in a liquid phase is
not necessarily related to the entropic component in the free energy
that has so far been assumed to be responsible for the formation of
equilibrium anisotropic liquid phases.

\subsection{Triply periodic fluid phase with $Fddd$ symmetry}\label{sec:fddd}

Bicontinuous partitioning of a volume is produced by inscribing a
continuous, orientable surface of positive genus without
self-intersection. This divides the volume into two multiply
connected, interpenetrating subvolumes, each of them physically
continuous (mathematically connected). Neither on a line nor in the
plane is there an analogue. A bicontinuous structure is a bicontinuous
partitioning in which each subvolume is filled with a distinct, not
necessarily uniform composition or state of matter. In familiar
examples one subvolume is solid or semi-solid, for example sandstone
or sponge. An interspersion of two phases is bicontinuous only if each
phase is connected across the specimen.  Scriven put forward the idea that bicontinuous structures may arise in
fluids.\cite{scriven}

\begin{figure}[htbp] %figure 3
\centering
\includegraphics[width=.7\columnwidth]{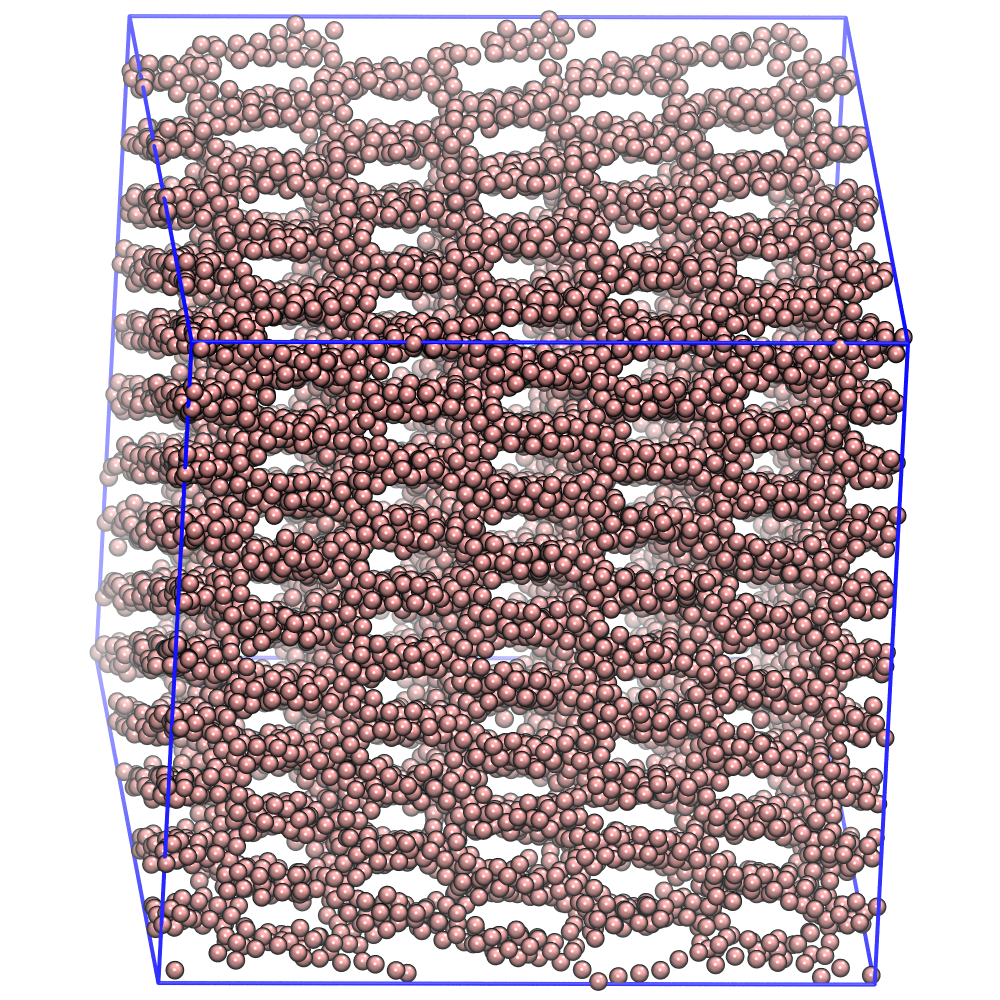}
\caption{ The instantaneous particle configuration of the simulated liquid
  mesophase with $Fddd$ symmetry. Reproduced from Ref.~\onlinecite{agosta_self-assembly_2020}, with the permission of AIP Publishing.}
\label{fddd_config}
\end{figure}
\begin{figure*}[tbp]
\includegraphics[width=\textwidth]{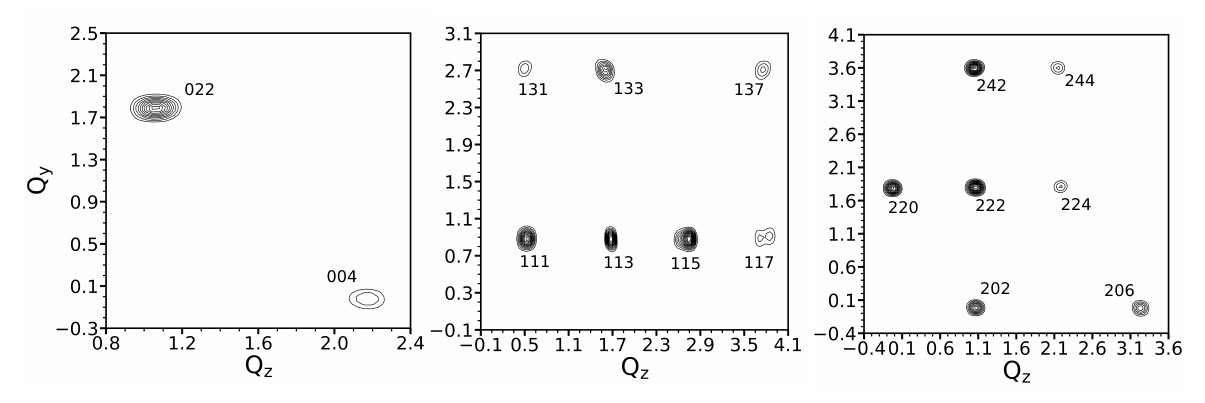}
\caption{ Isointensity contour plots of the $S({\bf Q})$ maxima labeled by
  Miller indices. Reproduced from Ref.~\onlinecite{agosta_self-assembly_2020}, with the permission of AIP Publishing.}
\label{fddd_sq}
\end{figure*}

The formation of triply periodic networks was observed in the systems
of block copolymers. In particle simulations they had so far only been
produced by anisometric particles, where the gyroid network forms
through a purely entropic mechanism,\cite{ellison_entropy-driven_2006,schonhofer_purely_2017}
or in binary mixtures of spherical particles with frustrated
cross-attraction,\cite{kumar_self-assembly_2017} in which the gyroid was
also found to be by far the most difficult mesophase to nucleate from
the disordered liquid.\cite{kumar_molinero_2018} For one-component
SALR fluids the double gyroid was shown to be thermodynamically stable
both by density-functional theory\cite{edelmann_gyroid_2016} and by
free-energy calculations in Monte Carlo
simulations;\cite{zhuang_prl_2016} in the latter, however, the network
could only be formed under an external modulated field imposing its
symmetry, and it did not self-assemble spontaneously from the
liquid. Inverse-designed isotropic potentials were found to form, at
high void fractions, a bicontinuous gyroid-like porous
fluid,\cite{lindquist_assembly_2016} but without the long-range
periodic order of a triply periodic network. The attempts to produce
the unassisted self-assembly of these morphologies in simulations of
one-component simple liquids were thus unsuccessful.

We describe here a simulation of a simple one-component system using the
SALR potential $V_4$ (System~II of
Ref.~\onlinecite{agosta_self-assembly_2020}) that produced
a spontaneous formation of a triply periodic liquid structure with the
$Fddd$ orthorhombic symmetry (Fig.~\ref{fddd_config}).

Triply periodic mesophases with the $Fddd$ orthorhombic symmetry were
found in block copolymers by Bailey et al.,\cite{bailey2002} opening
questions about the symmetry breaking factors that promote such
symmetry over the more entropically favoured cubic one; the family of
network phases accessible to block polymers continues to grow through
blending and through the design of low-$\chi$
triblocks.\cite{park_bates_2022,chang_grubbs_2017} In simple
systems of identical particles with spherically symmetric interaction
such transitions were never observed, although predicted by Landau
theories \cite{ciach_origin_2013,edelmann_gyroid_2016}.  The result we
discuss here is that such a mesophase self-assembles through a first-order
transition from the isotropic fluid phase. Interestingly, the
mesophase exhibits the same diffusion rate as the isotropic phase
while maintaining a complex 3D network at the mesoscale. The fluid
behaviour of triply periodic continuous networks represents a major challenge experimentally and it was
only hypothesized \cite{bailey2002}, as the standard diffraction
techniques can only determine a disordered internal structure and not
its fluidity.  For block copolymers it is conjectured
\cite{bailey2002} that the $Fddd$ morphology could arise from the
compromise between the enthalpic contribution of the single
molecular constituents and an entropic contribution arising from the
packing frustration directly related to the anisotropic shape of the
polymeric molecules.

\begin{figure}[htbp] %figure 3
\centering
\includegraphics[width=.62\columnwidth]{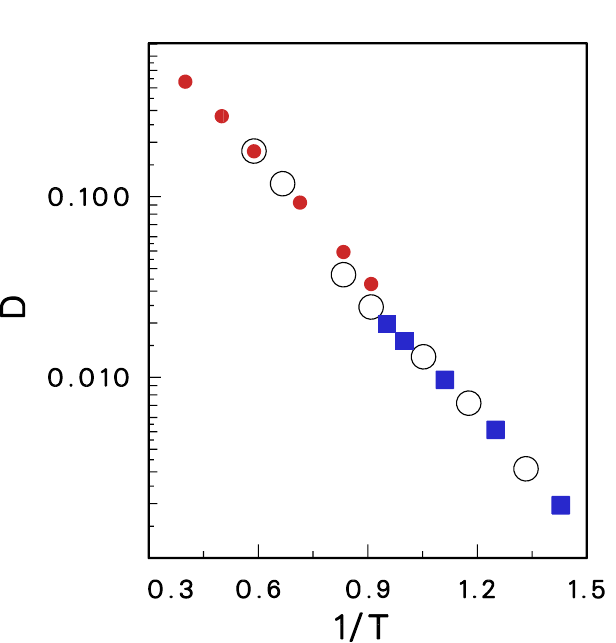}
\caption{ Arrhenius plot. Dots: isotropic liquid; squares:
Fddd phase, cooling; open circles: Fddd phase, reheating. Reproduced from Ref.~\onlinecite{agosta_self-assembly_2020}, with the permission of AIP Publishing.}
\label{fddd_arrh}
\end{figure}
\begin{figure*}[htbp]
\centering
\includegraphics[height=0.24\textheight]{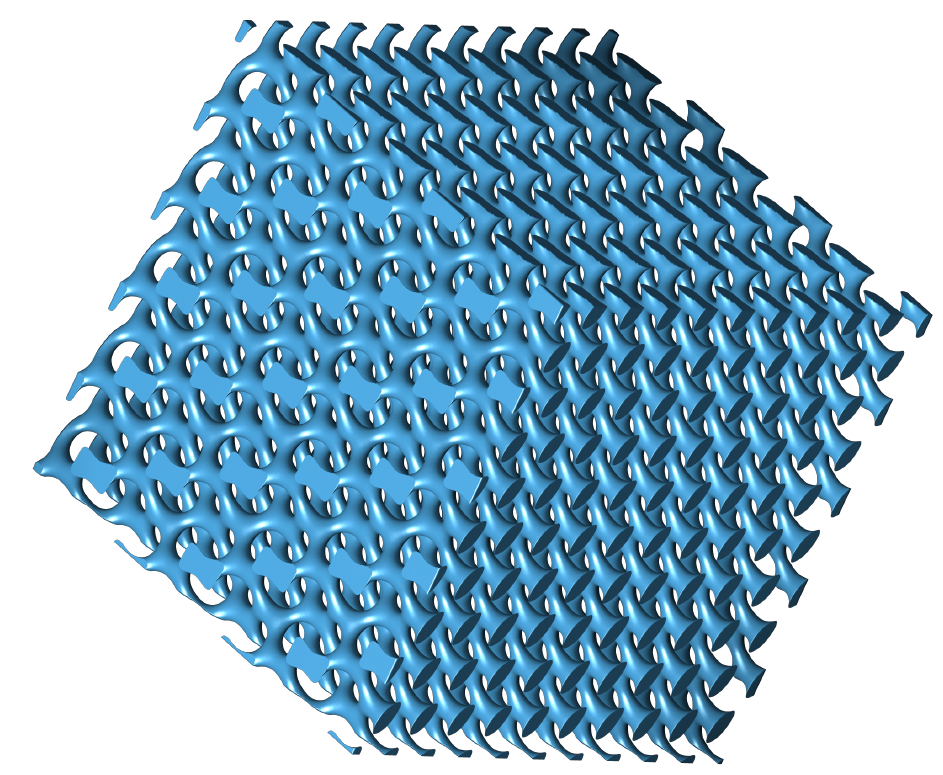}\hfill
\includegraphics[height=0.22\textheight]{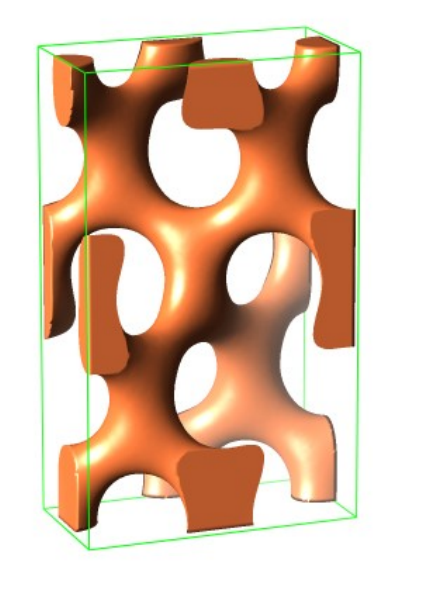}\hfill
\includegraphics[height=0.22\textheight]{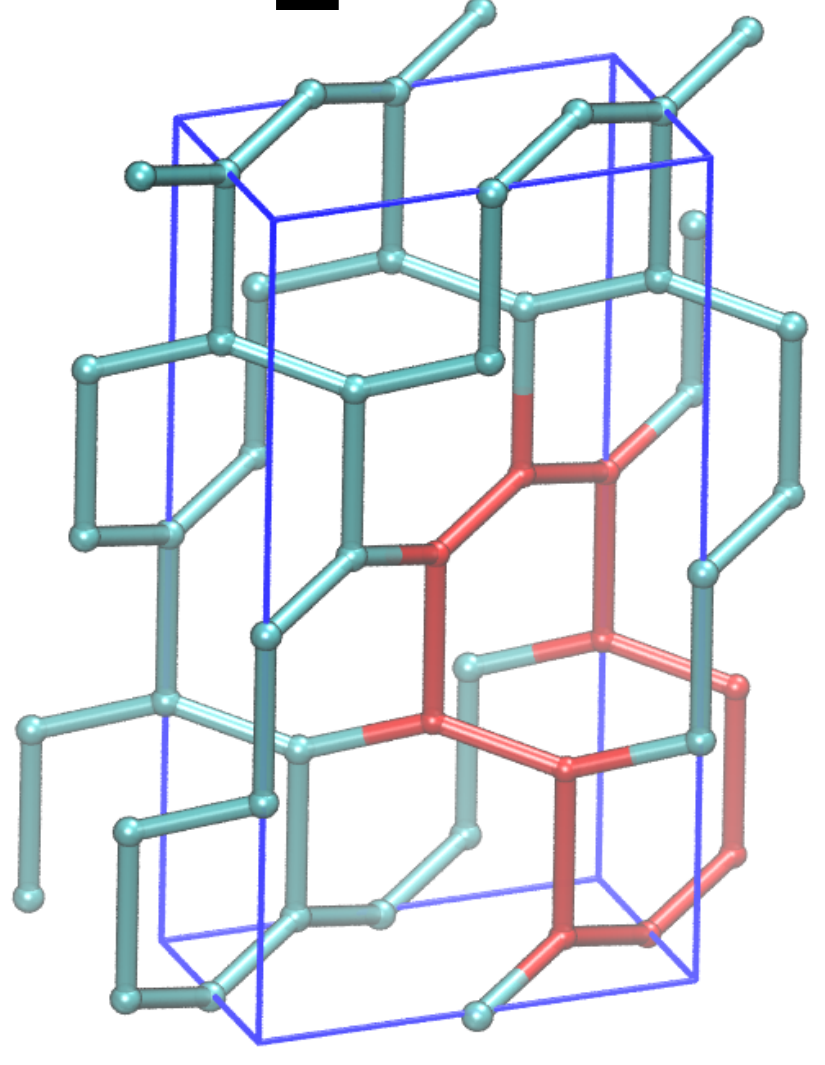}
\caption{ Left: the density isosurface calculated for the particle
  configuration of Fig.~\ref{fddd_config}. Centre: the unit cell cut from that
  density isosurface. Right: the ball-and-stick model, with the $(10,3)$ ring
  highlighted. Adapted from Ref.~\onlinecite{agosta_self-assembly_2020}, with the permission of AIP Publishing.}
\label{fddd_surf}
\end{figure*}

Figure~\ref{fddd_sq} shows the maxima of $S({\bf Q})$ in three characteristic
Q-planes labeled by Miller indices $hkl$. These patterns are consistent
with the small-angle x-ray scattering measurements on the $Fddd$
networks in block copolymers\cite{meuler_ordered_2009,jung}. This is the first observation of the
3D diffraction pattern of an $Fddd$ network.  The positions of the
observed maxima of $S(Q)$ are related to the orthorhombic unit cell
parameters $a, b, c$ as $Q_{hkl} = 2\pi(h/a, k/b, l/c)$. We found the
respective parameters as $(a:b:c) = (1:1.99:3.32), a = 3.49$. The ratios
of parameters reproduce those found in the $Fddd$
structures formed by block copolymers\cite{Epps2004,jung} within the margin of
the accuracy of measurements.  The $Fddd$ unit cell
parameters are intimately related to the problem of thermodynamic
stability of this orthorhombic mesophase relative to the cubically
symmetric gyroids. This symmetry breaking can be understood in terms
of weak segregation theory,\cite{leibler_theory_1980} which is the
block-copolymer form of the Landau argument put forward by Brazovskii,
Alexander and McTague for the selection of a modulated structure by
the superposition of density
modes.\cite{brazovskii_1975,alexander_mctague_1978} The Landau expansion of the free energy
in terms of composition modes demonstrates spinodal instability for
the modes corresponding to the lattice parameters obeying the
following relation: $(a:b:c) = (1:2:2\sqrt{3})$\cite{tyler}. The present lattice
parameters agree with this relation quite closely.

The diffusion of the constituent polymer chains along the separating
surface in the triply periodic networks formed by block copolymers is
blocked by the strong separation effects\cite{meuler_ordered_2009,Epps2004}. By contrast, the
simulated $Fddd$ morphology demonstrates liquid-like diffusion, see
Fig.~\ref{fddd_arrh}, which shows the Arrhenius plot of the system diffusion produced by heating and
cooling across the domain boundary. Neither the diffusion rate nor the
activation energy change upon the transition between the uniform
liquid and the $Fddd$ mesophase, demonstrating the fluid nature of the
latter. Thus, this transition is a microphase separation within the
liquid--gas spinodal domain, as conjectured in
Ref.~\onlinecite{lebowitz_rigorous_1966}.

Bicontinuous periodic morphologies can be conveniently presented
in terms of the triply periodic surfaces separating the two percolating domains. For the present morphology this is the density isosurface. A continuous, perfectly periodic real-space
density distribution has been produced by the inverse Fourier transformation of the calculated structure factor, the latter being
constrained by a Gaussian window function of an appropriately chosen
width. In this way a smooth density isosurface, free of the local density fluctuations, was produced; it is shown in Fig.~\ref{fddd_surf}, which also shows the unit cell of the simulated triply periodic
network as a fragment cut from the density isosurface.

The topology of triply periodic networks is commonly analysed in terms
of skeletal graphs and ball-and-stick models\cite{jung}. Such a model for the
unit cell is shown in Fig.~\ref{fddd_surf}. The stick lengths and
positions have been produced as the best fits to the isosurface.
According to the classification of Wells\cite{wells}, triply periodic nets can
be viewed as the tilings of space by closed loops of $n$ nodes with $p$
connections each, denoted as $(n, p)$. The $Fddd$ nets were concluded to
be composed of $(10,3)$ loops\cite{meuler_ordered_2009}. Such a loop is highlighted in the
ball-and-stick model shown in Fig.~\ref{fddd_surf}. We note
that the distance between the longer sticks coincides with the
parameter $a$ of the unit cell.  This distance is
consistent with the long repulsion range of the SALR potential.
Thus, we can rationalize the observed self-assembly of the $Fddd$
network in a simple system of particles as follows: the attraction
part of the pair potential favours condensation, whereas the range of
its long repulsion part defines the basic periodicity of the network,
and its symmetry arises from the Landau free energy minimum as a
result of the superposition of composition
modes\cite{leibler_theory_1980,tyler}.

\subsection{Other liquid periodic morphologies}

In Fig. \ref{phasediagram} the phase diagram for the potential $V_5$ for the number density interval 0.2-0.65 is reported. This potential form is able to capture all the standard mesophases typical of liquid crystals and block copolymers, such as micellar, columnar, triply periodic
structures, lamellar-smectic and inverse columnar phases.  We note
that the same phases can be obtained with the potential $V_4$ where
the differences between the generated structures lie in the
characteristic length scales of the formed mesophases. This makes it possible to tune the potential parameters to obtain the desired shapes
and geometries of the periodic phases.  We will now discuss the
main features of these mesophases.

For number densities lower than 0.24 spontaneous formation of micelle
lattices occurs. The formation of such a phase at low densities is not
surprising and has been reported previously for SALR
fluids.\cite{sciorti2004,zhuang_recent_2016} The same phase has been formed using
the potential $V_4$, which is characterised by a smaller size of
the micelles.  The micelles show internal liquid-like dynamics but
negligible particle exchange between micelles. The transition from
the bulk liquid is a first order transition.

For the densities between 0.25 and 0.38 columnar phases are formed. As
previously reported for the simulations using the potential $V_3$ the columns are arranged
in hexagonal order and are characterised by liquid diffusion only along
the axis parallel to the columns \cite{metere_formation_2015}. The
density variation affects the column diameter but not inter-column
distance; the latter is controlled by the long-repulsion part of the
potential.

Triply periodic continuous networks with $Fddd$ symmetry
appear in a small but well-defined region of the phase diagram. A similar orthorhombic phase arises in block copolymers.\cite{bates_network_2005} This is the same mesophase as that obtained with the potential $V_4$
and discussed in detail in Sec.~\ref{sec:fddd}, with a different unit
cell; its real-space surface model is shown in Fig.~\ref{fddd_surf}.

Finally at higher number densities we obtained lamellar phases characterised by having liquid diffusion along the plane, as in the smectic case, and the inverse columnar phases, where the particles diffuse like a liquid but are excluded from the empty columnar hexagonally packed vacuum regions (Fig.~\ref{phasediagram}).

\section{Discussion and perspectives}

\subsection{General comments}

A most general conclusion that can be drawn from the overview of the
wide range of simulated mesophases discussed here is that
particle-based computer simulations of simple one-component systems
using SALR-type pair potentials can be regarded as a very powerful
source of information, providing a unique insight into the microscopic
mechanisms of formation and thermodynamic stability of the mesophases
generated by the microphase separation transition as found in real
experiments. We showed that these simulations can also be used for
predicting new phases that have never been observed, or even those
which have never been thought of by theoreticians. It is remarkable
indeed that simulations of a simple system of identical particles
interacting via pair potentials of a rather unsophisticated functional
form can reproduce the structures of a wide variety of highly complex
and multicomponent mesophases arising in experiments on microphase
separation in real systems. Since the mesophases typical of block
copolymers and liquid crystals can be formed by a simple monatomic
liquid, the complexity of the constituent particles of real liquids,
whether their anisometric shape, their internal structure or their
chemical heterogeneity, is in principle not a necessary condition for
these phases to be generated. In this way, the present results can
rightfully be regarded as another demonstration of the success of the
central paradigm of physics that can be formulated as reductionism:
the belief that the apparent complexity of the observable natural
phenomena can be reduced to simple models formulated in terms of a few
fundamental principles.

\begin{figure*}[tbp]
\includegraphics[width=1.0\textwidth]{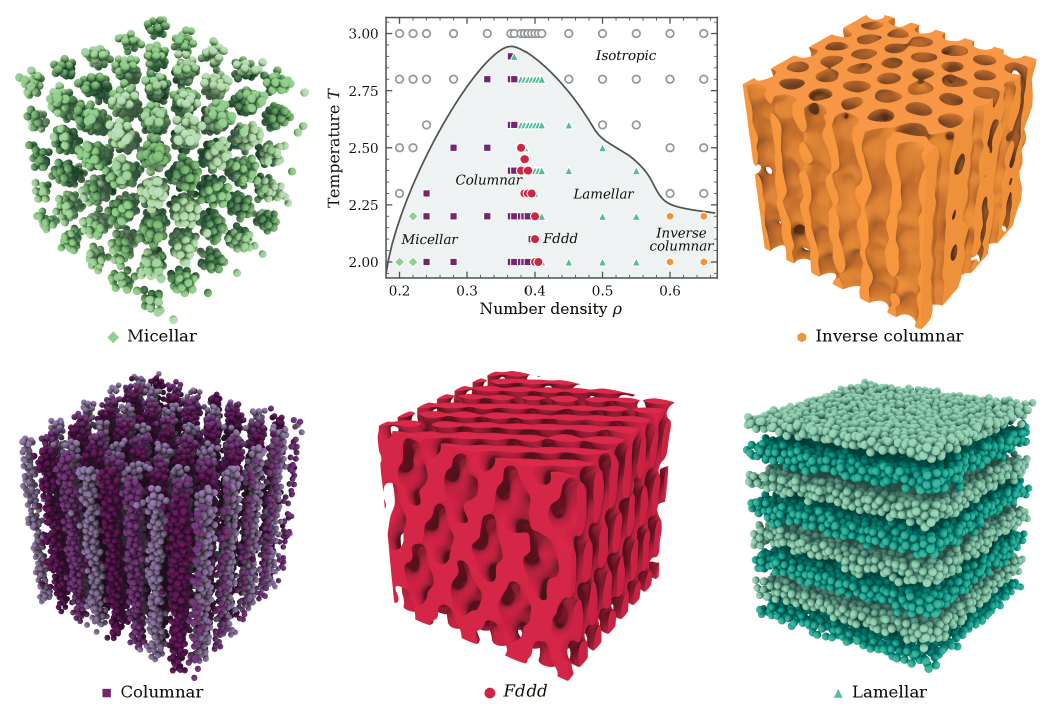}
\caption{Temperature--density phase diagram of the system with the
  potential $V_5$ (top centre), surrounded by snapshots of the five
  mesophases taken from molecular-dynamics configurations of
  $N=16\,384$ particles. Open grey circles: isotropic liquid. Green
  diamonds: micellar phase. Purple squares: columnar phase. Red
  circles: $Fddd$. Teal triangles: lamellar phase. Orange hexagons:
  inverse columnar phase. The solid line is a guide to the eye marking
  the boundary of the isotropic liquid. Each snapshot is a cubic block
  cut from the periodic configuration and oriented along the symmetry
  axes obtained from its structure factor. Micellar: every micelle is
  drawn in a single shade, viewed along the [111] direction of the bcc
  micellar lattice. Columnar: the columns are vertical, with three
  shades distinguishing neighbouring columns of the hexagonal
  lattice. Lamellar: alternate layers in two shades. $Fddd$ and inverse
  columnar: isosurface of the Gaussian-smoothed particle density at the
  mean density, showing the continuous network and the hexagonal array
  of empty channels, respectively.}
\label{phasediagram}
\end{figure*}

\subsection{Physical properties and technological applications of mesophases}
The discovery of a class of pair potentials able to generate a rich
variety of solid and fluid periodic mesophases opens the possibility
to address conceptually important problems in the physics of soft
matter. In particular, it makes it possible to accurately sample the
relevant phase diagrams to identify critical points, triple and
tri-critical points,\cite{Varta1,elenius_dzugutov_2009} or the proximity of the area of the
phase diagram where strong density fluctuations from the isotropic to
the one-dimensionally periodic meso-structure can be observed. Another
interesting aspect is the investigation of the isotropic phase prior
to microphase separation. Such a liquid behaves like a foam when
approaching the spinodal domain, probably indicating the existence of
a liquid-liquid phase transition. The possibility to resolve the
particle dynamics in such complex geometries is fundamental for measuring characteristic time correlation functions that could help
experimentalists to characterise real structures.

The realisation of the described mesophases could be directly
achieved in real systems of particles interacting with the class of
SALR potentials reported so far. We note that in the reported
meso-structures the first minimum of the SALR potential (short
attraction) always corresponds to the packing distance of the single
particles, while the range of the SALR shoulder (long repulsion)
defines the meso-scale length (the distance between micelles, columns,
planes, or the pores of the 3D networks). Good candidates are colloidal particles whose interaction can in principle be
modelled and tuned to achieve the tailored interaction, as described by the DLVO theory.\cite{VO,IS} Due to sluggish kinetics, experiments have so far only produced gels and cluster
phases\cite{sciorti2004,campbell_dynamical_2005}. However, it was shown recently that enclosing a colloidal fluid
with competing interactions in an ordered porous matrix of primitive,
diamond or gyroid symmetry can template ordered
microphases,\cite{serna_2020} provided that the template
is commensurate with the intrinsic periodicity of the fluid.
The inverse route, in which a target structure is specified and the
pair potential that stabilises it is sought numerically, has also been
demonstrated for SALR cluster
fluids\cite{lindquist_assembly_2016,lindquist_inverse_2021} and could be applied directly to the functional form of Eq.~(\ref{potential}).

Soft-matter systems whose constituents are themselves spherical on the
mesoscale are natural candidates for the smectic dodecagonal
quasicrystal of Sec.~\ref{sec:qc}. Dodecagonal order has been produced by
dendrimers,\cite{ZENG} star
polymers\cite{DO2,ren_star_2016} and binary nanoparticle
superlattices,\cite{TAL} and globular protein--polymer conjugates have
been shown to form the lamellar, hexagonal and gyroid mesophases of
block copolymers;\cite{lam_olsen_2013,lam_olsen_2014} in all these
systems the effective interaction between the mesoscopic units is, to
a first approximation, spherically symmetric.
Ionic liquid crystals are a second class of candidates: their
smectic phases arise from the microphase segregation of the ionic and
the alkyl parts of the molecules,\cite{goossens_2016,saielli_2015}
i.e. from a competition between a short-range attraction and a
longer-range electrostatic repulsion that is the molecular analogue of
the SALR potential. Recently, the formation of an $Fddd$ network with liquid
diffusion was observed in a solid-state ionic
conductor\cite{famprikis_new_2019} at high temperature. This opens the
possibility to investigate this class of materials to find the right
mix of ionic components able to stabilise triply periodic mesophases
at ambient conditions. Hybrid organic--inorganic perovskites belong
in the same category: the dimensionality of their inorganic framework,
from three-dimensional to layered two-dimensional and one-dimensional
phases, is selected by the organic cation, and we have recently shown
that non-covalent interactions between the cations, such as chalcogen
bonding and S--$\pi$ interactions, can be used to direct the assembly
of low-dimensional phases with improved conductivity and
stability.\cite{luo_advsci_2024} The transport properties of SALR
fluids, which determine whether a mesophase can be reached on the
experimental time scale, have themselves become a subject of
systematic study.\cite{perdomo_2022}

Liquid metals offer a further, less obvious connection. The structure
factors of many liquid alloys, among them the Al--transition-metal,
Al--rare-earth and Li--Pb systems, exhibit a prepeak at wave vectors
below the main peak, the signature of a structural correlation on a
scale of several atomic diameters. The origin of this
intermediate-range order has been debated since it was first
catalogued:\cite{price_1989} it has been attributed to chemical
short-range order, to the packing of transition-metal-centred
polytetrahedral clusters\cite{roik_2010} and, in network liquids and
glasses, to the ordering of voids.\cite{elliott_1991} Simulations of
a simple monatomic liquid interacting through the pair potential from
which the SALR form of Eq.~(\ref{potential}) was derived showed that
such a prepeak can arise in a one-component system from the ordering
of vacancies on a medium-range scale.\cite{dzugutov_mro_1998,sadigh_1999}
In SALR fluids the same low-$Q$ feature, there called the
intermediate-range-order peak, is the recognised diagnostic of
incipient microphase separation and of the cluster, columnar and
lamellar states that follow from
it.\cite{godfrin_2013,bollinger_fluids_2016,liu_colloidal_2019} We
therefore conjecture that the prepeak of liquid alloys may be the
precursor of a liquid mesophase of the kind described in this review,
with the effective interatomic interaction, whose Friedel oscillations
provide a natural source of a long-range repulsion,\cite{hafner_1987,elenius_dzugutov_2009,elenius_oppelstrup_dzugutov_2010}
playing the role of the SALR potential. A direct test would be the
detection, in the reciprocal-space pattern of the metallic liquid, of
the symmetry of a periodic modulation at the wave vector of the
prepeak.

\subsection{Metamaterials}

Metamaterials are topological materials, defined at the mesoscopic
scale, and produced in a property-oriented-design
philosophy.\cite{kadic_3d_2019} Among them, materials built on triply
periodic minimal surfaces and related geometries occupy a special
place, because the same few surfaces (primitive, diamond, gyroid)
recur from biological tissue and self-assembled soft matter to
engineered lattices,\cite{han_overview_2018} and because their
mechanical, transport and optical properties can be tuned through
the surface topology and the relative density alone.\cite{alketan_2019}
The scale is typically of the order of several times the size of the constituent molecules (100~nm--10~$\mu$m). This implies that locally these phases are disordered, like amorphous or liquid structures, and that their formation and stability depend strictly on the thermodynamic conditions of the environment.

Three-dimensional periodicity at this scale can result in the enhancement of
physical properties independently of the chemical composition;
importantly, this effect is scale invariant. Hence the goal is to create a specific, material-independent topology, which is what links a given 3D network to the desired physical properties.

For instance the same topology can be exploited, depending on the
typical size of the problem, for photonic band gaps and topological
photonic states (100--400~nm),\cite{maldovan_photonic_2002,optical_meso,ling_lu_weyl_2013,VARD}
for gas storage and separation and for energy storage at the
nanometre scale, where three-dimensionally continuous porous graphene
with gyroid-like topology combines a very large accessible surface
with mechanical robustness and electrical
continuity,\cite{han_graphene_2022,qin_buehler_2017,yang_branicio_2026}
for micro mixing and chromatography
(10--1000~$\mu$m),\cite{morton_hydrodynamic_2008,biagioni_2025} and for
acoustic wave propagation (mm), mechanical stress and thermal
conductivity/isolation (cm)
properties.\cite{alketan_2019,meuler_ordered_2009,li_bioinspired_2020}
The chromatographic application deserves a specific remark, since it
illustrates how the \emph{geometry} of a periodic structure, rather
than its material, controls a transport property. It was recently
shown that a liquid-chromatography column whose channel is partitioned
by a periodic minimal (helicoidal) surface reduces the minimal plate
height by a factor of two to three with respect to the best packed,
monolithic and micro-pillar columns, without any penalty in pressure
drop, because the minimal surface maximises transversal mixing of the
mobile phase while minimising the friction
losses.\cite{biagioni_2025} Triply periodic networks of the kind
produced in the simulations reviewed here are the natural
three-dimensional generalisation of that idea.

There is a subclass of MOF structures, named meta-MOFs, which share
many important exotic features with metamaterials. The term was
introduced by Coudert and Evans\cite{coudert_metamofs_2019} for
crystalline framework materials whose anomalous behaviour, such as
negative thermal expansion, negative linear or area compressibility,
auxeticity and pressure-induced softening, originates not from the
chemistry of the building units but from the geometry of their periodic
arrangement, in direct analogy with the wine-rack, honeycomb and hinged
motifs of macroscopic mechanical metamaterials; the analogy has recently
been made literal by the design of an origami-like MOF that reproduces
the folding mechanics of a mechanical
metamaterial.\cite{jin_origami_mof_2023} As for metamaterials, the
property of a meta-MOF is thus a property of its net topology rather
than of its chemical composition. In Sec.~\ref{sec:mof} we described the
self-assembly of a structure with MOF topology from a simple
one-component liquid simulated using a SALR potential. This suggests
the perspective that new kinds of meta-MOF topologies can self-assemble
in these systems, selected by the thermodynamic state of the fluid
and by the range of the long repulsion rather than by chemical design,
and that their mechanical response can be assessed directly in the
simulation, as was done above for the layer-gliding anisotropy of the
smectic phases (Fig.~\ref{qc_shear}).

A related development is the recent introduction of
\emph{spinodoid} metamaterials,\cite{kumar_kochmann_2020} non-periodic
cellular topologies that approximate the microstructures produced by
spinodal decomposition and whose property space can be explored and
inverse-designed with a handful of parameters. The microphases
reviewed here stand in the same relation to spinodal decomposition as
the SALR potential stands to the van der Waals model: the long
repulsion arrests the coarsening at a prescribed length and replaces
the random bicontinuous pattern by a periodic one. SALR simulations
therefore provide the periodic counterpart of spinodoid design, with
the symmetry of the resulting network selected by the thermodynamic
state rather than by construction.

\subsection{Experimental and computational challenges}

Shaping colloidal particles which would self-assemble in a desired
geometry requires long experimental trials to find the right
colloidal interactions. Moreover, the possibility of tuning triply
periodic networks with non-trivial 3D surfaces at the mesoscale is still
far from being achieved.  Liquid crystals and block-copolymers are two
classes of materials that can form mesophases \cite{bates_network_2005} and
then be used for producing metamaterials. However the formation of periodic 3D topologies at large scales useful for applications is very
challenging. To remedy this problem many attempts have been made using colloidal particles. Colloids are in principle easy to manipulate and
they can be designed to have specific pair interactions that will
result in formation of super structures like
mesophases.\cite{DERJAGUIN199330} Some improvement has been obtained by creating colloidal particles with anisotropic shapes that are able to stick to each other and pack in ordered structures. This approach produced some 3D structures at the mesoscale, such as the diamond lattice, which were shown to give enhanced photonic properties.\cite{diamond}
Nevertheless, these types of assemblies can only generate trivial
topological structures, as the result of the predetermined particle
anisotropy.  Triply periodic phases, like gyroids, are not
accessible by this experimental methodology, due to their topological
complexity. Advances have been reported in 3D printing techniques of
gyroids and other structures with a known analytical surface
description, both in polymers, where polymerisation-induced phase
separation now gives printed objects an inherent nanoporosity on the
sub-micrometre scale,\cite{dong_levkin_2021} and in inorganic
materials such as transparent silica glass printed at low
temperature.\cite{li_glass_2023} This solution is particularly
attractive as analytical surfaces are scale-free and can be 3D printed
for a wide range of applications. The $Fddd$ network is the case in
point where this route is blocked: no analytical minimal surface with
that symmetry is known, so the numerical isosurface produced by the
simulation of Sec.~\ref{sec:fddd} is at present the only available description of
its geometry.

A route that circumvents both the kinetic hindrance of colloidal
self-assembly and the need for an analytical surface description is
the inverse templating of the self-assembly of block copolymers (or of
lyotropic surfactants): the triply periodic morphology formed by the
soft matter is used as a mould, one of the blocks is removed, the
continuous void is back-filled with an inorganic material and the
polymer scaffold is finally eliminated. In this way the gyroid double
network of the surfactant-templated silica MCM-48 and of the
block-copolymer-templated large-pore silica KIT-6 was replicated into
continuous carbon and platinum networks,\cite{kleitz_kit6_2003} and the
three-dimensional structures of these replicas and of their parent
silicas were solved by Terasaki and co-workers by electron
crystallography.\cite{sakamoto_nature_2000,kaneda_jpcb_2002,sakamoto_angew_2004}
The same strategy applied to triblock terpolymer gyroids has produced
mesoporous titania electrodes for hybrid solar cells,\cite{crossland_2009}
gyroidal superconductors,\cite{robbins_2016} gold gyroid optical
metamaterials with a unit cell of about 50~nm,\cite{vignolini_2012,stefik_2015}
and, very recently, a mesoporous gyroidal strontium titanate
perovskite.\cite{moore_srtio3_2025} Since the replica inherits the
symmetry and the characteristic length of the template, the phase
diagrams of the SALR systems reviewed here, which contain the same
lamellar, hexagonal and network morphologies with a period set by the
range of the long repulsion, identify the thermodynamic conditions under
which a given template, including the $Fddd$ network, is expected to
form.

On the computational side, molecular dynamics simulation of mesophase
formation can be used as a guidance for the production of self-assembled structures like naturally formed liquid crystals and
polymers. These simulations
are normally coarse-grained and use particles that imitate the shape
of the real organic mesomorphic molecules\cite{glotzer2007} or their
competitive interactions (two or more components)
\cite{kumar_self-assembly_2017}, thereby forming similar final self-assembled structures. Because of the dimensions of the organic
constituents, the accessible system size that can be reached by such
MD simulations is limited, and extended bulk phases can hardly be
generated. As shown in this Perspective, the introduction of
spherically shaped pair potentials that are expected to approximate the effective interaction of larger molecules is the key to simulate
mesophase formation for large length and time scales. In particular
SALR potentials\cite{zhuang_recent_2016,lindquist_assembly_2016}, if
properly tuned, can produce equilibrium structures typical of polymers
and liquid crystals (see Fig.~\ref{phasediagram}). Before the
simulations reviewed here, 3D mesophases of SALR fluids could only be
stabilised by an external field imposing their symmetry (Sec.~\ref{sec:fddd}). The difficulty is
kinetic as much as thermodynamic: even in binary mixtures that do
form the gyroid at equilibrium, its nucleation from the disordered
liquid is orders of magnitude slower than that of the lamellar and
hexagonal phases.\cite{kumar_molinero_2018} The direct, unassisted
self-assembly of the $Fddd$ network reported in Sec.~\ref{sec:fddd} shows that
with an appropriately shaped pair potential this difficulty can be
overcome in a one-component system.

The molecular dynamics simulations could then be
exploited to generate a class of mesophases of the desired shapes,
symmetries and characteristic lengths oriented to specific physical
properties. It is possible to transform an equilibrium mesophase
produced in a simulation using a system of particles into a surface
network that can eventually be produced in material form using the 3D
printing technologies. This would establish the ultimate link between
simulations and real structures bringing a revolutionary change in the
way of fabricating and modelling metamaterials.

\begin{acknowledgments}
We thank Prof.\ Osamu Terasaki, Prof.\ Sven Lidin, Prof.\ Peter
Oleynikov, Prof.\ Michael O'Keeffe, Dr.\ Tomas Oppelstrup and Dr.\ Sten
Sarman for their help in understanding and characterising the phases
discussed here.
\end{acknowledgments}

\section*{Author declarations}
\subsection*{Conflict of interest}
The authors have no conflicts to disclose.

\section*{Data availability}
The data that support the findings of this study are available
from the corresponding author upon reasonable request.

\bibliography{references,extra_refs}% Produces the bibliography via BibTeX.

\end{document}